\documentclass[usenatbib]{mn2e} 

\usepackage{graphicx}
\usepackage{anysize}
\usepackage{multirow}
\usepackage{epstopdf}
\usepackage{natbib}
\marginsize{2cm}{2cm}{1cm}{2cm} 

\newcommand{\kms}{km\,s$^{-1}$}

\newcommand{\lala}{$\lambda$$\lambda$}
\newcommand{\ergs}{erg\,s$^{-1}$}

\newcommand{\Hb}{H{\sc$\beta$}}

\newcommand{\cii}{C\,{\sc ii}}
\newcommand{\ciiid}{C\,{\sc iii}]}
\newcommand{\ciis}{C\,{\sc ii}$^{*}$}
\newcommand{\civ}{C\,{\sc iv}}

\newcommand{\nv}{N\,{\sc v}}

\newcommand{\siii}{Si\,{\sc ii}}
\newcommand{\siiis}{Si\,{\sc ii}$^{*}$}
\newcommand{\siiv}{Si\,{\sc iv}}

\newcommand{\niii}{Ni\,{\sc ii}}
\newcommand{\niv}{Ni\,{\sc v}}

\newcommand{\alii}{Al\,{\sc ii}}
\newcommand{\aliii}{Al\,{\sc iii}}

\newcommand{\feii}{Fe\,{\sc ii}}
\newcommand{\feiis}{Fe\,{\sc ii}$^{*}$}
\newcommand{\feiii}{Fe\,{\sc iii}}
\newcommand{\feiiis}{Fe\,{\sc iii}$^{*}$}

\newcommand{\crii}{Cr\,{\sc ii}}
\newcommand{\znii}{Zn\,{\sc ii}}

\newcommand{\mgii}{Mg\,{\sc ii}}

\title[Vanishing Absorption Lines in FeLoBAL Quasars]{Vanishing Absorption and Blueshifted Emission in FeLoBAL Quasars}

\author[A. Rafiee et al.]{Alireza Rafiee,$^{1}$\thanks{E-mail: arafiee@yorku.ca}
Patrik Pirkola,$^{1}$
Patrick B. Hall,$^{1}$
Natalee Galati,$^{1}$
\newauthor
Jesse Rogerson,$^{1}$
Abtin Ameri$^{1,2}$\\
$^{1}$Department of Physics and Astronomy,
York University, 4700 Keele St., Toronto, ON M3J 1P3;\\
$^{2}$Bloor Collegiate Institute, 1141 Bloor St. West, Toronto, ON M6H 1M9.
}

\begin{document}
\label{firstpage}
\pagerange{\pageref{firstpage}--\pageref{lastpage}}
\maketitle

\begin{abstract}
We study the dramatic decrease in iron absorption strength in the iron low-ionization broad absorption line quasar SDSS J084133.15+200525.8.
We report on the continued weakening of absorption in the prototype of this class of variable broad absorption line quasar, FBQS J140806.2+305448.
We also report a third example of this class, SDSS J123103.70+392903.6; unlike the other two examples, it has undergone an increase in observed continuum brightness (at 3000~\AA\ rest-frame) as well as a decrease in iron absorption strength.
These changes could be caused by absorber transverse motion or by ionization variability.
We note that the \mgii\ and UV \feii\ lines in several FeLoBAL quasars are blueshifted by thousands of \kms\ relative to the \Hb\ emission line peak.  We suggest that such emission arises in the outflowing winds normally seen only in absorption.
\end{abstract}


\section{Introduction}\label{intro}

Quasars are a subtype of active galactic nuclei (AGN). A quasar consists of a super-massive black hole (SMBH) surrounded by an accretion disk \citep{ss73} which can emit enough light to outshine the entire surrounding galaxy.
Viscosity within the accretion disk converts gravitational potential energy to thermal energy \citep{net13}. The sum of black-body curves from the disk leads to a power law spectrum in the rest-frame near-UV and optical \citep{kn99}.

Scattering or absorption of flux by surrounding media results in broad or narrow absorption line troughs in quasar spectra.
A broad absorption line (BAL) quasar is, historically, defined as a quasar that exhibits blueshifted absorption due to the \civ\ doublet at \lala\ $1548.203$, $1550.770$~\AA\ that is at least $2000$~\kms\ wide and is found in the velocity range $3000$~\kms\ to $25000$~\kms\, where $0$~\kms\ is at the systemic redshift of the quasar and positive velocities indicate motion toward the observer. 

As discussed in \cite{sdss123}, there are three types of BAL quasars. High-Ionization BALs exhibit absorption due to relatively high-ionization ions such as \civ, \siiv, or \nv.
Low-Ionization BALs (LoBALs) also exhibit broad absorption due to
elements in comparitively lower ionization states, such as \aliii\ or \mgii. If the quasars also exhibit broad absorption due to excited states or levels of \feii\ or \feiii, they are referred to as Iron LoBALs (FeLoBALs) ~\citep{sdss123}.

Approximately $25$\%\ of quasars exhibit BAL phenomena (see discussions in \citealt{RH11} and \citealt{AH11}), and this fraction increases
when narrower ($500-2,000$~\kms) mini-BAL troughs are included in the census
(see \citealt{RHH11} for a full discussion on mini-BAL quasars).

In the context of the disk-wind model of luminous AGN, broad and blueshifted absorption is interpreted as obscuring material lifted off the accretion disk surrounding the SMBH and accelerated by radiation line driving and perhaps magnetohydrodynamical effects to high outflow velocities (e.g., \citealt{MC95}, \citealt{OC10}).
The disk-wind model is widely accepted as it explains, in broad terms,
the observed features in quasar spectra; however, it does not constitute a
fully formed theory of accretion physics and outflows.

Outflows, manifested as BALs, provide insight into the dynamical and chemical properties of the central engines of quasars (e.g., \citealt{elv00}, \citealt{AC15}, and see \S~4.5 of \citealt{DB10}).  Moreover, at such high outflow velocities and sufficiently large kinetic luminosities, BAL winds may also represent a mechanism by which SMBHs provide feedback to their host galaxy
and play a significant role in massive galaxy formation and evolution (e.g.,
\citealt{MA09},
\citealt{AB13},
\citealt{LT14},
\citealt{CA15},
\citealt{hopkins15}).

Variability of broad absorption troughs has been widely reported
in the literature as far back as the early 1990s
(e.g., \citealt{GB08}, \citealt{fbqs1408}; see Table~1 in \citealt{FB13}
for sample of BAL variability studies).
These works find that BAL troughs vary in their depth, width, and velocity
profile on both short and long timescales. The cause of variability
is likely due to a mixture of transverse motion of
absorbing clouds across our line of sight (e.g., \citealt{fbqs1408}) and
changes in the ionization of the absorbing gas or changes in the optical depth of the absorbing gas (e.g., \citealt{HK08},
\citealt{FB13}). By studying the variability of these dynamic troughs,
we can gain new insight into the central engines of AGN,
the physical parameters of their outflows, and the connections between
those outflows and the AGN host galaxies.

In this paper we present spectroscopic 
and photometric data 
on three quasars that appear to form a sub-class of quasars transitioning away from FeLoBAL status through loss of iron absorption.
We identify and discuss troughs in which absorption has changed, present the spectral energy distributions (SEDs) of all three quasars, and discuss the implications of these observations.
We adopt cosmological parameters of $H_0=70$ km~s$^{-1}$~Mpc$^{-1}$, $\Omega_M=0.3$ and $\Omega_\Lambda=0.7$.
\section{Spectroscopic and Photometric Data} \label{Phot}
The SDSS (Sloan Digital Sky Survey) began operations in 2000. The survey used a dedicated 2.5m telescope with CCD imaging in the $ugriz$ bands \citep{doi10} and spectroscopy to catalog objects over a large area of the sky \citep{yor00}. The original spectrograph produced over 1.5 million spectra, and as part of SDSS-III \citep{sdss3} was upgraded in 2009 for the BOSS (Baryon Oscillation Spectroscopic Survey) \citep{bosssmee}. To date the SDSS has produced spectra for over 300,000 quasars (Paris et al.\ 2015, in preparation).  The BOSS took spectra over $\sim$3600$-$10400~\AA\ in the observed frame, expanding on the original survey range ($\sim$3800$-$9200~\AA).
%
The other astronomical surveys used for data analysis in this paper come from CRTS, GALEX, 2MASS, and WISE.  CRTS (Catalina Real-Time Transient Survey) is a synoptic astronomical exploration covering 33,000 square degrees of the sky \citep{crts}; the Catalina Sky Survey (CSS) component of the CRTS obtained the data used in this paper.  GALEX (Galaxy Evolution Explorer) is a space telescope collecting data in the ultraviolet range \citep{galex}. 2MASS (Two Micron All Sky Survey) scanned the sky in the near-infrared J, H and K bands \citep{2MASS}. WISE (Wide-Field Infrared Survey Explorer) collected data at four infrared wavelengths centred between 33 and 221 $\mu$m \citep{wise}.

\subsection{Quasar Redshifts}
\label{redshift}

For convenience, we give the redshifts we adopt for our quasars here, before discussing each quasar individually.

For the quasar SDSS J084133.15+200525.8 (hereafter J0841), we adopt a redshift of $z=2.337\pm0.003$ based on matching the non-BAL composite quasar of \cite{sdssbal} to the spectrum of J0841 at rest-frame $>$2700\,\AA.
The SDSS DR10 quasar catalog \citep{bossdr10q} gives a visual inspection redshift of $z=2.342$.

For the quasar SDSS J123103.70+392903.6 (hereafter J1231), we adopt the Principal Component Analysis redshift $z_{PCA} = 1.0227 \pm 0.0004$ from the SDSS DR12 quasar catalog (DR12Q, Paris et al.\ 2015, in preparation).
Note that this differs from the $z=1.004$ reported in passing by \cite{2015ApJS..217...11L} and the $z=1.0038$ adopted by \cite{mcgraw+2015}.

For the quasar FBQS J140806.2+305448 (hereafter J1408), we adopt the value of $z=0.848$ used in \cite{fbqs1408}, for consistency.
The SDSS spectrum redshift according to \cite{hw10} is $z=0.845\pm 0.001$, while the BOSS spectra yield a lower redshift of $z_{PCA}=0.8310\pm 0.0004$ (Paris et al.\ 2015, in preparation).
These different redshifts are likely related to the emission-line blueshifts discussed in \S~\ref{bshifts}.

\subsection{Quasar J0841}

\begin{figure}
\centering
\includegraphics[width=\columnwidth]{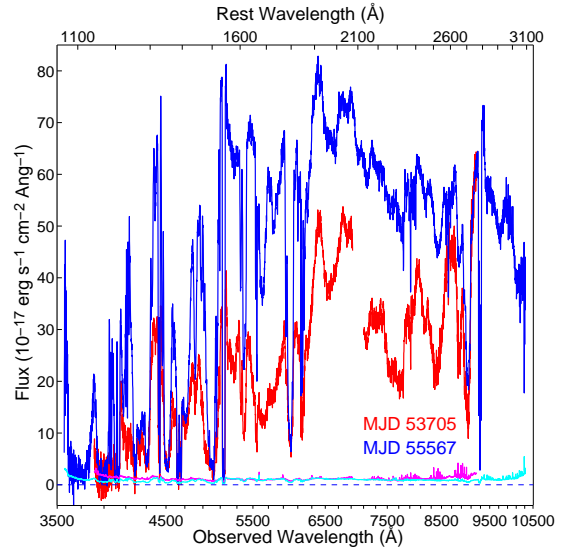}
\caption{Spectra of J0841 in the SDSS spectral epoch (red) and the BOSS epoch (blue). The magenta and cyan curves on the botton correspond to the 1-$\sigma$ uncertainties of the flux of the red and blue epochs respectively. The bottom axis shows observed-frame wavelengths, and the top axis shows rest-frame wavelengths. The top axis is labeled using a redshift of z = 2.337. The tick marks of the top axis are every 100 \AA $\:$with labels every 500 \AA. The tick marks of the bottom axis (facing out) are every 200 \AA $\:$with labels every 1000 \AA. The BOSS epoch is normalized as a function of wavelength, with the condition that the differences between $9095 < \lambda < 9210$~\AA\ are minimized (see \S \ref{spec}).}\label{fig:j0841}
\end{figure}

\subsubsection{Spectrophotometric Correction}\label{spec}

We use improved-sky-subtraction DR7 spectra from \cite{wh10}.

The BOSS spectrophotometric standard stars were taken through fibers placed to optimize throughput at 5400~\AA.  However, the BOSS spectrum of J0841 was taken through a fiber placed to optimize throughput at 4000~\AA\ \citep{bossover}.  Thus, the spectrophotometric flux calibration of the BOSS spectrum of J0841 is systematically in error.
\cite{tpcorr} have modeled this effect and calculated a wavelength-dependent correction factor $C_\lambda$ for each affected BOSS spectrum.

In the case of J0841, we find that the correction predicts too large a change in brightness as compared to that observed in the CRTS observations (see \S \ref{HistPhot}).
We therefore rescale the correction $C_\lambda$ to $C'_\lambda=1+B(C_\lambda-1)$ such that the value of $C'_\lambda$ at $9095 < \lambda < 9210$~\AA\ brings the SDSS spectrum into agreement with BOSS spectrum at those wavelengths.
Furthermore, using this rescaled correction instead of the original Margala et al.\ correction brings the magnitude differences between the SDSS and BOSS epochs estimated from  spectroscopy into better agreement with the magnitude difference between those epochs measured using CRTS photometry.  (Because CRTS uses a broad, non-standard filter only approximately calibrated to the $V$ band, we do not attempt to compare magnitudes.  Instead, we compare just the change in the CRTS $V$ magnitude to the changes in our synthesized $g$ and $r$ magnitudes.)

We check our rescaled correction using nine other quasars with spectra on the same two plates as J0841.  A rescaled correction with the same value of $B$ is a better fit in six of nine cases (assuming the only difference between epochs for each quasar was due to the need for correction).
One possible explanation for why the rescaled correction is a better fit is that the seeing when our spectra were taken was on average better than assumed by the correction algorithm.

In Figure \ref{fig:j0841}, we plot the SDSS and rescaled BOSS spectra of J0841, from epochs 53705 and 55567 (2005 and 2011) respectively. We observe a large change in flux due to an absorption decrease between the two spectra.

\subsubsection{Black Hole Mass Estimate}

For J0841, the value of $L_{bol}$ from 	\cite{2011ApJS..194...45S} is suspect because it is based on a partial detection  of \mgii\ in the SDSS spectrum, whereas the BOSS spectrum covers the whole \mgii\ region and the luminosity at $2900$~\AA\ as suggested by \mgii\ calibrations, see e.g., \cite{2011ApJS..194...42R}, \cite{Assef+2011}.

Using $F_{2900}=5.5\times 10^{-16}$ erg s$^{-1}$ cm$^{-2}$ \AA$^{-1}$,
$D_L=5.8\times 10^{28}$ cm, 
and a 2900 \AA\ bolometric correction of $BC_{2900}=5\pm 1$ \citep{gtr06},
we find
$L_{bol}=(BC_\lambda)\lambda F_\lambda 4\pi D_L^2= (3.36\pm 0.69) \times 10^{47}$ \ergs.


Due to strong absorption within \mgii\ emission line, a clean black hole mass measurement is not possible from this object's BOSS spectrum.
For our calculations, we adopt
$M_{BH} = 6 \times 10^9 ~M_\odot$
and thus $L_{Edd}=3.8\times 10^{47}$ erg s$^{-1}$ 
with Eddington ratio of $L_{bol}/L_{Edd} = 0.45$. 
The black hole mass and Eddington ratio are used to calculate transverse velocities of bulk motion in our discussion.

\subsubsection{Trough Identification}

\begin{figure*} 
\centering
\includegraphics[width=13cm,height=5cm]{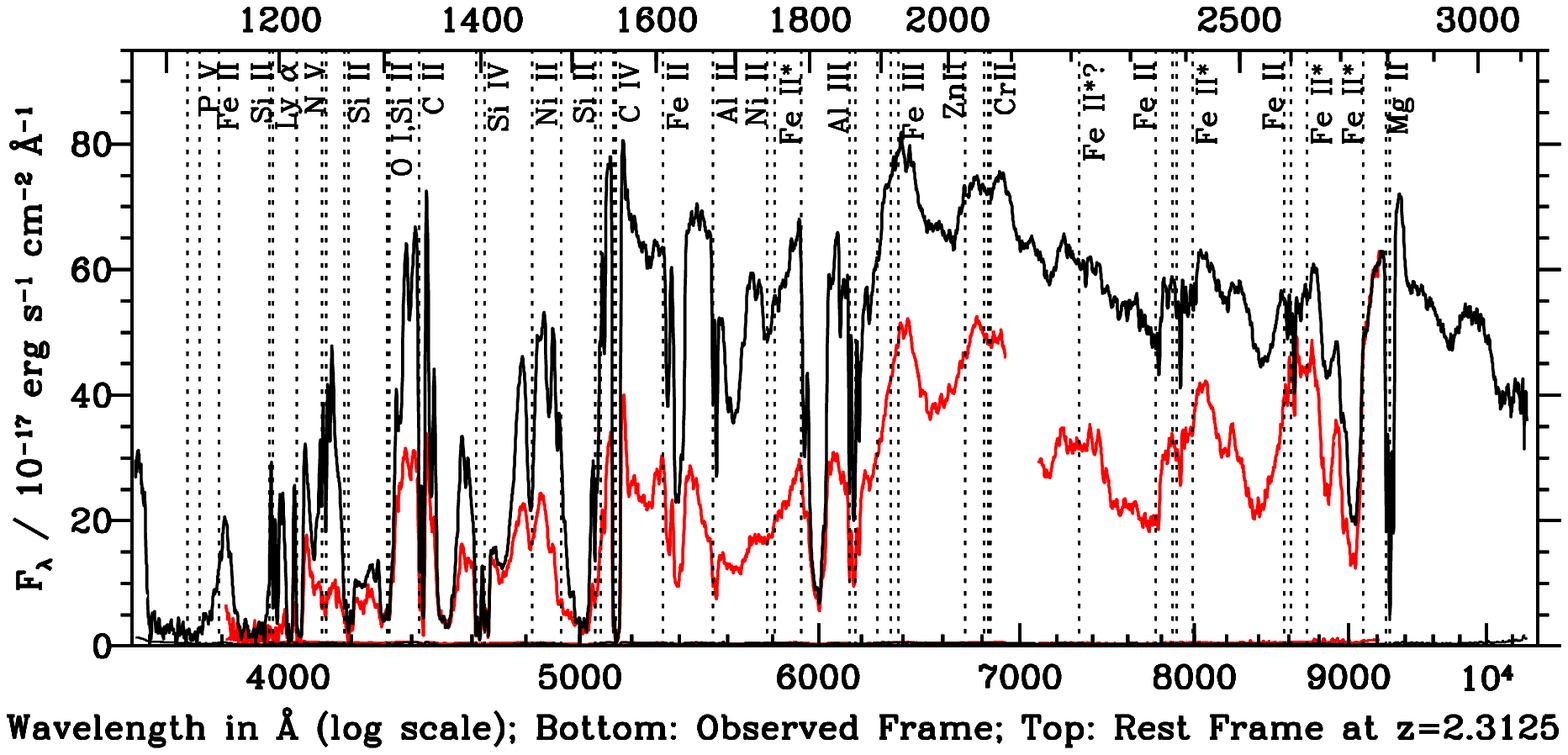}
\includegraphics[width=13cm,height=5cm]{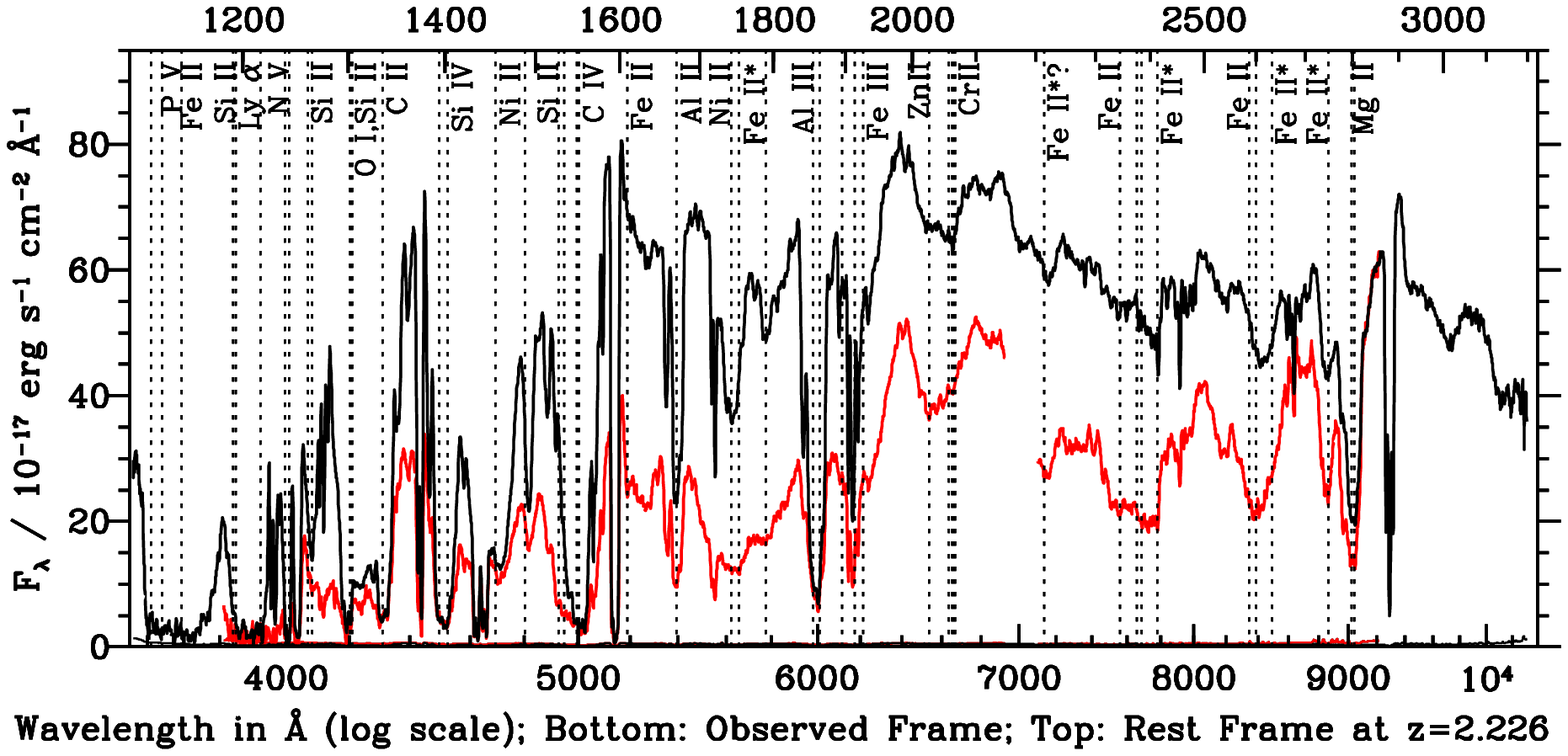}
\vspace{0.2cm}
\caption{\label{fig:f_j0841n}Spectra of J0841 in the SDSS spectral epoch (red) and the BOSS epoch (black).  In each panel the bottom axis shows observed-frame wavelengths, and the top axis shows rest-frame wavelengths.  Notable absorption lines are labeled along the top, and are marked by dotted vertical lines.  Top: Absorption in the rest frame at $z=2.3125$, the frame of the highest-redshift of three closely spaced associated absorption systems.  Not all the absorption lines listed are present at this redshift.  Bottom: Absorption in the rest frame at $z=2.226$, at which broad absorption lines are strongest.  All absorption lines listed are believed to be present at this redshift.}
\end{figure*}

In this quasar there is blended narrow associated absorption
at $z=2.3225$, $z=2.3184$, and $z=2.3125$,
or line-of-sight blueshifts of 1310 \kms, 1680 \kms, and 2210 \kms\ relative to the adopted rest frame of $z=2.337$ (see Fig. \ref{fig:f_j08magvsmjd}).
The absorption at $z=2.3225$ and $z=3.3125$ is visible in many transitions,
but with relatively weak \feii\ and possibly \feiis\ absorption.
The normalized residual flux does not change significantly between the two epochs in the transitions with the deepest troughs but does increase in transitions with less deep troughs, such as \aliii.
The absorption at $z=2.3184$ is only seen directly in \nv\ and \aliii, which have the widest doublet separations among detected transitions;
other transitions are too blended for this system to be distinguished.
(An intervening \mgii\ absorption system at $z=0.8166$ produces a narrow absorption doublet near 5090 \AA.)

The broad absorption is strongest at $z=2.226$,
or a line-of-sight blueshift of 10,100 \kms.
Centred on that redshift, there are overlapping troughs of \feii\
(from both ground and excited states)
which weaken dramatically between our two spectral epochs.
Also weakening are troughs of
\siii\ $+$ \siiis, \niii, \alii, \feiiis, \znii, \crii, and \mgii.

However, the bottoms of the troughs in
\niv, \cii\ $+$ \ciis, \civ\ and \aliii\ are consistent with having the same flux at both epochs.
Unchanged flux in both epochs likely indicates saturated absorption in those troughs.

The \civ\ absorption trough is $\geq$10,000 \kms\ wide,
with line-of-sight blueshifts spanning 4,300$-$14,400 \kms.
The \aliii\ and \mgii\ troughs are $\geq$4000 \kms\ wide.

In the second spectral epoch, the typically strong \civ\ emission line
is beginning to be recognizable, along with the 1900~\AA\ emission
complex (consisting of \ciiid, \feiii, and other lines), the
2080~\AA\ emission complex of \feiii, the \mgii\ emission line, and
the characteristic shape of \feii\ emission complexes longward of \mgii.

The small fractional increases in flux near 4250 \AA\ and 4700 \AA\ between our two spectra are at odds with what is seen at nearby wavelengths away from strong, saturated troughs.
One possible explanation for the lack of change at these wavelengths is nearly saturated high-velocity \civ\ and \siiv\ absorption $z=2.035$ (28,400 \kms\ outflow velocity).
Another possible explanation would be high initial optical depths in \siii\ and perhaps \niii, coupled with overlapping absorption from \feii\ and other transitions affecting those wavelengths.

Troughs which do not vary between the two epochs consist of
absorption which is saturated in both epochs.
The residual flux level in such a trough decreases with increasing covering factor of the gas, although the value of the covering factor cannot be established without knowing the continuum level.
However, we can see that the narrow \civ\ systems have higher covering factors than the broad \civ\ trough.
The covering factor is also higher for saturated troughs at shorter wavelengths, consistent with a continuum source size which is smaller at shorter wavelengths, due to short-wavelength emission being produced in high temperature regions preferentially located relatively close to the black hole.

\subsubsection{Equivalent Width}
\label{EWj0841}
The equivalent width difference between the two epochs is calculated in two ways: (1) with respect to a straight line connecting the two ends of the \mgii, \feii\ or combined \mgii$+$\feii\ troughs \citep{fbqs1408}, (2) with respect to a non-BAL quasar composite fit to the later epoch at wavelengths longer than 9500 \AA\ \citep{sdssbalcat}. Since the estimates are very similar we only report the linear estimate in Table \ref{tab:DEW1}. The equivalent width of  \feii\ absorption shows a dramatic decreases of  about $49.59\pm 0.83$~\AA\ while \mgii\ absorption shows a decrease, during the same time interval, of about $29.02\pm 1.00$~\AA, see column 2 in Table \ref{tab:DEW1}.


\subsubsection{Historical Photometry}\label{HistPhot}

\begin{figure} 
\centering
\includegraphics[width=\columnwidth]{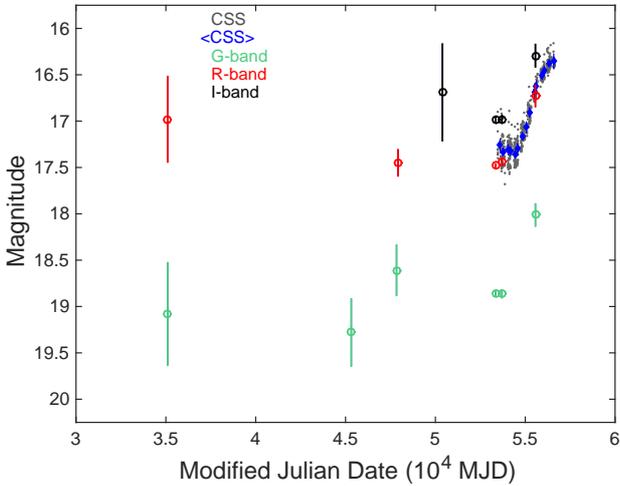}
\caption{\label{fig:f_j08magvsmjd}Above is the light curve of quasar J0841. The Modified Julian Dates (MJDs) are tabulated in Table \ref{tab:tphot}. The data are grouped in the $g$, $r$, and $i$ bands indicated by the colors on the graph, as well as in the optical band provided by CSS (Catalina Sky Survey). $<$CSS$>$ refers to the average value of CSS data discussed below.}
\end{figure}


\begin{table*}
\footnotesize
\caption{Photometry of SDSS J084133.15+200525.8}
\label{tab:tphot}
\begin{tabular}{|cccccccc|}
\hline
Date       & MJD      & Source  & Original         & Syn.           & $g_{est}$        & $r_{est}$        & $i_{est}$     \\ \hline 
1954-12-20 & 35097.41 & POSS-I  & O=19.78$\pm$0.52 & 19.56$\pm$0.17 &  19.08$\pm$0.55  & ...              & ...            \\ 
1954-12-20 & 35097.42 & POSS-I  & E=16.65$\pm$0.43 & 17.17$\pm$0.15 & ...              &  16.98$\pm$0.46  & ...            \\
1982-12-16 & 45319.96 & Palomar & V=18.42$\pm$0.36 & 18.12$\pm$0.05 &  19.28$\pm$0.36  & ...              & ...            \\
1989-11-07 & 47838.49 & POSS-II & J=17.79$\pm$0.25 & 18.04$\pm$0.09 &  18.61$\pm$0.27  & ...              & ...            \\
1990-01-27 & 47919.28 & POSS-II & F=17.35$\pm$0.07 & 17.42$\pm$0.12 & ...              &  17.45$\pm$0.14  & ...            \\ 
1996-11-04 & 50391.99 & POSS-II & N=16.87$\pm$0.49 & 17.16$\pm$0.16 & ...              & ...              &  16.69$\pm$0.52  \\
2004-12-12 & 53351.40 & SDSS    & $g,r,i$ PSF      & ...            & 18.856$\pm$0.020 & 17.481$\pm$0.015 & 16.980$\pm$0.024 \\ 
2005-12-01 & 53705.97 & SDSS    & spectrum         & ...            & 18.866$\pm$0.038 & 17.434$\pm$0.027 & 16.979$\pm$0.038 \\ 
2011-01-04 & 55565.82 & BOSS    & corr.\ spectrum  & ...            & 18.012$\pm$0.118 & 16.729$\pm$0.113 & 16.294$\pm$0.120 \\ 
\hline
\end{tabular}
\end{table*}
\normalsize

J0841 is bright enough to have been detected in photographic plates in the last century.
To search for variability over decadal timescales,
we compared our SDSS magnitudes to measurements from the Palomar Sky Surveys
(POSS-I, Palomar Quick V, and POSS-II).
The results are shown in Table \ref{tab:tphot}.
Magnitudes labeled Original in the Table are the averages of the magnitudes
from the USNO-B \citep{usnob1} and GSC2.3.2 \citep{2008AJ....136..735L}, when available,
with the exception of the POSS-II J and F epochs.
For those epochs, we use the re-calibrated magnitudes of \cite{sesar06}.
The synthesized magnitudes (labeled Syn.\ in the Table) are what the historical surveys would have seen if this object had the same flux then as it did during the SDSS imaging epoch.
Synthesized magnitudes are computed from the observed SDSS magnitudes using
\cite{usnob1} equations (2a) and (2b) for the POSS-I epochs,
\cite{sesar06} equation (3) and Table \ref{tab:tphot} for the POSS-II epochs,
and the equation of \cite{2005AJ....130..873J}
for the Palomar Quick V epoch.
For the latter we adopt a nominal uncertainty of 0.20 magnitudes, while for the other synthesized magnitudes we adopt uncertainties equal to the quoted scatter in the relevant conversion equation.

To aid in searching for long term trends, for each epoch we also give an estimated magnitude in the SDSS band closest to the observed photometric band. 
For example, for photometric bands similar to the $g$ band, we calculate $g_{est} = {\rm ~Original~} - {\rm ~Syn.~} + g_{SDSS}$.
In other words, we translate the difference between the observed and synthesized magnitudes to the closest SDSS band, yielding a simplistic but useful estimate of the SDSS magnitude at different epochs.

Comparing the original and synthesized magnitudes at each epoch, and the estimated photometry in each individual filter over time, we see there is no evidence for a significant change in the magnitudes of J0841 prior to the SDSS imaging epoch.

For the two spectral epochs in Table \ref{tab:tphot} (SDSS and BOSS),
the magnitudes listed are calculated from the {\tt spectroFlux} values for each filter via
$ \rm mag = 22.5 - 2.5 \times \log_{10}({\tt spectroFlux}) $.
The uncertainties are the {\tt gRMSStd}, {\tt rRMSStd}, and {\tt iRMSStd} values, which are already in magnitudes.

\indent
The data in Figure \ref{fig:f_j08magvsmjd} was retrieved from the sources tabulated in Table \ref{tab:tphot} and from the CRTS. The g-band represents wavelengths in the visual range; the r-band lies just outside the range of visual in the infrared; the i-band has a wavelength in the near infrared. Refer to Table \ref{tab:tphot4} for more wavelengths in the different bands. These data points were chosen for a range of MJDs specified from Table \ref{tab:tphot}. The g-band has 6 data points, the r-band has 5 data points and the i-band has 4 data points; despite the difference in the number of points, the relationship between the magnitude and Modified Julian Dates (MJDs) for the different bands is very similar. Focusing on MJDs between 47000$-$57000 where there are four data points for each filter, the trend is very similar. This and the fact that the change in the magnitudes of the individual bands are generally greater than the 2$\sigma$ error leads to the conclusion that these changes in magnitudes are most likely due to an intrinsic property of the object.

The changes in flux are caused primarily by BAL variability.
The continuum level is uncertain by $\pm$10\%, and so continuum
variability can contribute to the flux variability at that level, but
the flux varies by a much larger percentage.  The constant flux at
$\sim$4250~\AA\ observed rules out a substantial change in reddening
as a cause of the increasing flux.

CRTS was another dataset used, specifically the CSS telescope (refer to section \S~\ref{Phot} for more information on the survey). This data provides vital insights into the behaviour of the quasar. The grey data points in Figure \ref{fig:f_j08magvsmjd} represent the direct CRTS data and the blue points are the associated averaged values of this data. The average data was calculated for 15 bins, each with a size of approximately 42.5 data points in length. The average values demonstrate a tangent hyperbolic curve; the magnitude was level at a constant value, then had a polynomial change before leveling off again at another constant value. This change lies within a 2$\sigma$ error as seen on the graph, which again implies a likely explanation stemming from an intrinsic property of the quasar and not due to an observing error or otherwise. It should also be noted that when scaling the data in the other bands, they follow a similar trend and shape to that of the CRTS data  which provides further validity to the results.


\subsection{Quasar J1231} \label{j1231}

\begin{figure} 
\centering
\includegraphics[width=\columnwidth]{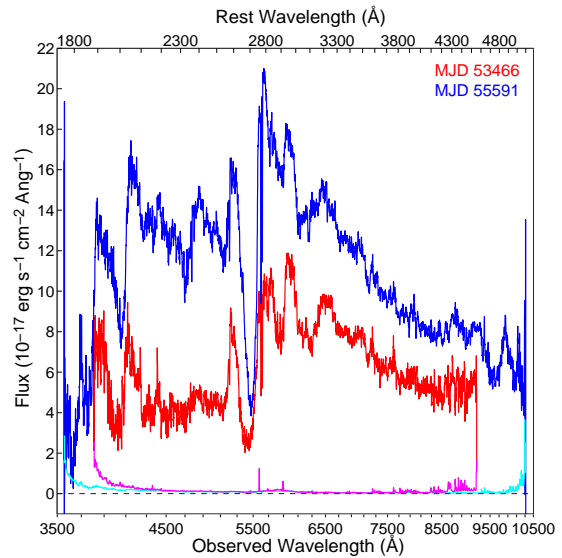}
\caption{\label{fig:f_j1231}Spectra of J1231. The magenta and cyan curves on the bottom correspond to the 1-$\sigma$ uncertainties of the flux in the red and blue epochs, respectively. The bottom axis shows observed-frame wavelengths, and the top axis shows rest-frame wavelengths at a redshift $z = 1.0227$. The tick marks inside the box are spaced every 100 \AA $\:$with labels every 500 \AA. The tick marks of the bottom axis (facing out) are spaced every 200 \AA $\:$with labels every 1000 \AA. The spectra are smoothed with a 5 pixel box-car weighted average.}
\end{figure}

\begin{figure} 
\centering
\includegraphics[width=\columnwidth]{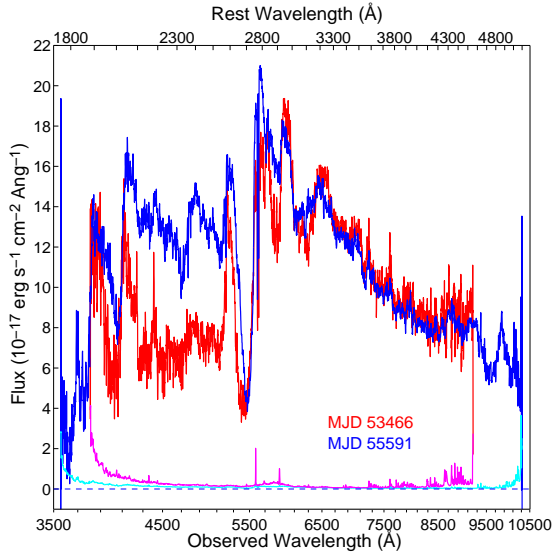}
\caption{\label{fig:f_j1231n}Spectra of J1231 normalized over 7200$-$8500 \AA\ observed frame. The magenta and cyan curves on the bottom correspond to the 1-$\sigma$ uncertainties of the red and blue epochs respectively. The bottom axis shows observed-frame wavelengths, and the top axis shows rest-frame wavelengths at a redshift $z = 1.0227$. The tick marks inside the box are spaced every 100 \AA $\:$with labels every 500 \AA. The tick marks of the bottom axis (facing out) are spaced every 200 \AA $\:$with labels every 1000 \AA. The spectra are smoothed with a 5 pixel box-car weighted average.}
\end{figure}

Figure \ref{fig:f_j1231} is a plot of the spectra of J1231 from MJDs 53466 and 55591; there is obvious strong flux variability. \cite{mcgraw+2015} also report this variability; however, they report a lack of detectable strong variability in other spectra taken by them within about one rest-frame year of the MJD 55591 spectrum.
Figure \ref{fig:f_j1231n} is a plot of the same spectra normalized at wavelengths $7200-8400$~\AA\ observed.

\subsubsection{Power Law Fitting and Analysis} \label{j1231plaw}

\begin{figure} 
\centering
\includegraphics[width=\columnwidth]{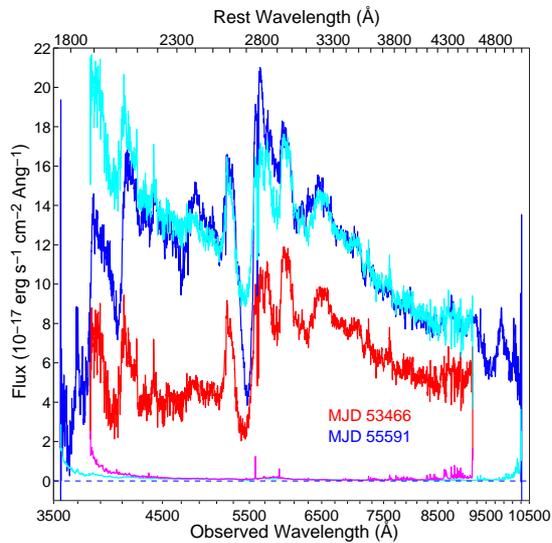}
\caption{\label{fig:f_j1231a}Spectra of J1231 and a simulated second-epoch spectrum (cyan; see \S~\ref{j1231plaw}). The red and blue curves on the bottom correspond to the 1-$\sigma$ uncertainties of the flux in the red and blue epochs, respectively.
}
\end{figure}
No spectrophotometric correction is needed for the BOSS spectra of this object, as its fiber was positioned to optimize throughput at the same central wavelength as for the standard stars.

To analyze absorption and continuum changes, we fit a power-law to this object's continuum.  Such a fit is reasonable due to the relatively absorption free regions at rest-frame $\lambda>3000$~\AA.

For each spectrum, we find the best-fit amplitude and slope of a power law continuum 
$F_\lambda=A(\lambda/3400.7$~\AA$)^\alpha$,
with $\lambda$ being the rest-frame wavelength in \AA.
The fitting regions used were $3021 \leq \lambda \leq 3100$~\AA, $4145 \leq \lambda \leq 4157$~\AA, and $4202 \leq \lambda \leq 4227$~\AA, which are all continuum regions in the non-BAL composite spectrum presented in \cite{sdss73}.
These regions were chosen to give a good range of wavelengths in the nearly absorption-free region $3021 < \lambda < 4250$~\AA. We note that the region $3021 \leq \lambda \leq 3100$~\AA\ may contain weak absorption, but there is no evidence for strong absorption there in the normalized spectra.
For J1231,
for the first epoch we found $\alpha=-1.6$ with
$\chi^{2}_{\nu}  \approx 0.787$ and for the second epoch $\alpha=-1.8$ with $\chi^{2}_{\nu}  \approx 1.05$.
We estimate the systematic uncertainty to be $\pm$0.2 for each index fit. These fits are used for black hole mass estimate (see next section) and simulation of quasar evolution (see next paragraph).

When our two spectra of J1231 are multiplicatively normalized by a constant factor equal to the ratio of their fluxes in the rest-frame 3600 - 4200~\AA\ region, the profile of the low-velocity half of the Mg\,{\sc ii} BAL is the same in both spectra, and the spectra longward of 3100~\AA\ are identical as well.
That indicates identical brightenings of regions not covered by the Mg\,{\sc ii} BAL (seen in the bottom of the BAL trough) and regions covered by it (which contribute most of the flux seen at wavelengths outside of absorption troughs).
If instead only a part of the source uncovered by Mg\,{\sc ii} had brightened, we would expect the second-epoch spectrum to be well fit by the first-epoch spectrum plus a smooth power-law component, which is not the case (see the cyan curve in Figure \ref{fig:f_j1231a}).  Similarly, if only a part of the source covered by Mg\,{\sc ii} had brightened, we would expect no change at Mg\,{\sc ii} and an increase in flux at other wavelengths, which is not the case either.

The different behaviours exhibited by the Fe\,{\sc ii} and Mg\,{\sc ii} troughs indicate that Fe\,{\sc ii} does not cover as much of the source as Mg\,{\sc ii}.
The possibility that the Mg\,{\sc ii} and Fe\,{\sc ii} covering is the same but the optical depth of Fe\,{\sc ii} is less than that of \mgii\ is less likely, because the shape of the Fe\,{\sc ii} troughs between 2200 and 2600~\AA\ in the first epoch indicate strong saturation \citep{sdss123}.

Brightening of the entire continuum source could occur either through a change in mass accretion rate $\dot{M}$ propagating through the disk, or a change in outer disk illumination due to a sustained brightening of the inner disk.  Evidence of outer disk illumination changes in response to emission changes close to a black hole has been seen by \cite{2015ApJ...806..129E}.

\subsubsection{Black Hole Mass Estimate} \label{j1231mbh}

Unlike \cite{mcgraw+2015}, for J1231 we do not use the $M_{BH}$ and $L_{bol}$ values from \cite{2011ApJS..194...45S} because they are based on incorrect fits to the \mgii\ region that ignore the absorption troughs.
To estimate the black hole mass of J1231, we use the rest-frame FWHM of the \Hb\ line and the flux at 5100 \AA\ from the power law fit of the previous section.
Using the suggested \Hb\ calibration of \cite{Assef+2011}, we find  $\log M_{BH} = 9.313 \pm 0.077$.
For that $M_{BH}$, we have $L_{Edd}=3.08\times 10^{47}$ erg s$^{-1}$
using Eq.\ 3.16 of \cite{net13}:
$L_{Edd}=(1.26\times 10^{38}{\rm erg~s}^{-1})(M/M_\odot)$, appropriate for solar metallicity gas.
Using $F_{5100}=5\times 10^{-17}$ erg s$^{-1}$ cm$^{-2}$ \AA$^{-1}$,
$D_L=2.1\times 10^{28}$ cm, 
and a 5100 \AA\ bolometric correction of $BC_{5100}=10.4\pm 2.5$ \citep{gtr06},
we find
$L_{bol}=(BC_\lambda)\lambda F_\lambda 4\pi D_L^2=(1.47\pm 0.35)\times 10^{46}$ erg s$^{-1}$.
That luminosity yields $L_{bol}/L_{Edd} = 0.058$. The black hole mass and Eddington ratio are used to calculate transverse velocities of bulk motion in our discussion.


\subsubsection{Trough Identification} \label{j1231ids}

\begin{figure*} 
\centering
\includegraphics[width=13cm,height=5cm]{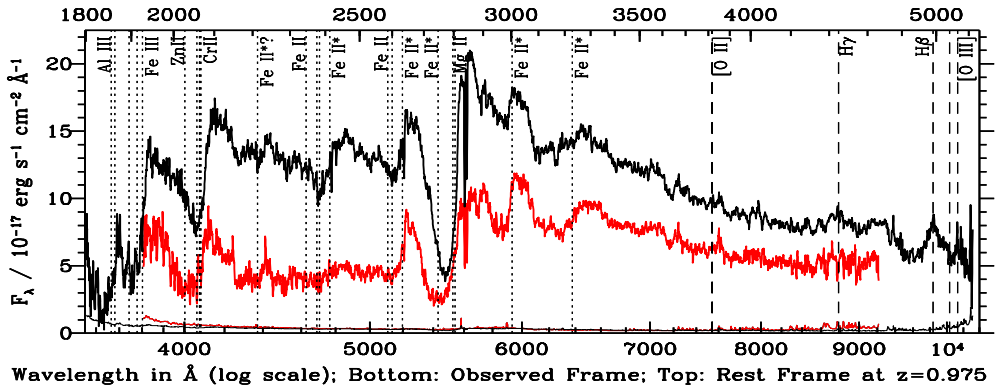}
\caption{\label{fig:f_j1231plog}
Spectra of J1231 in the SDSS epoch (red) and the BOSS epoch (black), with 1-$\sigma$ uncertainties plotted along the bottom in the same color. Notable absorption- and emission-line wavelengths are shown by dotted and dashed lines, respectively, and are labeled along the top.
In each panel the bottom axis shows observed-frame wavelengths, and the top axis
shows rest-frame wavelengths at $z=0.975$, the redshift of the onset of absorption troughs.  (Emission lines are plotted in the quasar rest frame of $z=1.0227$.)}
\end{figure*}

As seen in Figure \ref{fig:f_j1231plog}, J1231 shows broad absorption from \mgii, \feii, \feiis, blended \crii\ $+$ \znii, \feiii, and \aliii.\footnote{J1231 also shows narrow (FWHM $<$ 1000 km s$^{-1}$) absorption from
\mgii\ and \feii\
at $z=1.0034\pm 0.0001$, a blueshift of 2870 $\pm$ 60 km s$^{-1}$ from the adopted systemic redshift.}

Relative to the adopted systemic redshift of $z=1.0227$, the absorption trough
spans blueshifts of $7150-14830$ km s$^{-1}$ ($z=0.925$ to $z=0.975$).  There is an uncertainty of $\pm$160 km s$^{-1}$ ($\pm 0.01$ in $z$)
on the start velocity, which was determined by finding the wavelengths at which the troughs of the ions mentioned above had reached on average half their maximum depth (the troughs have very well-defined long-wavelength edges). The end velocity is determined by where the troughs appear to recover to a local continuum, but is more uncertain ($\pm$480 km s$^{-1}$, or $\pm 0.03$ in $z$) due to blending with other troughs.
The overall trough width is $7680 \pm 510$ km s$^{-1}$.

\subsubsection{Equivalent Width}
\label{EWj1231}
The equivalent width of \feii\ absorption shows a decrease of about $33.67\pm 1.51$~\AA\ while \mgii\ absorption decreases for about $14.44\pm 1.55$~\AA\ between two epochs, see column 3 in Table \ref{tab:DEW1}.

\subsubsection{Historical Photometry} \label{histphot}

\begin{figure} 
\centering
\includegraphics[width=\columnwidth]{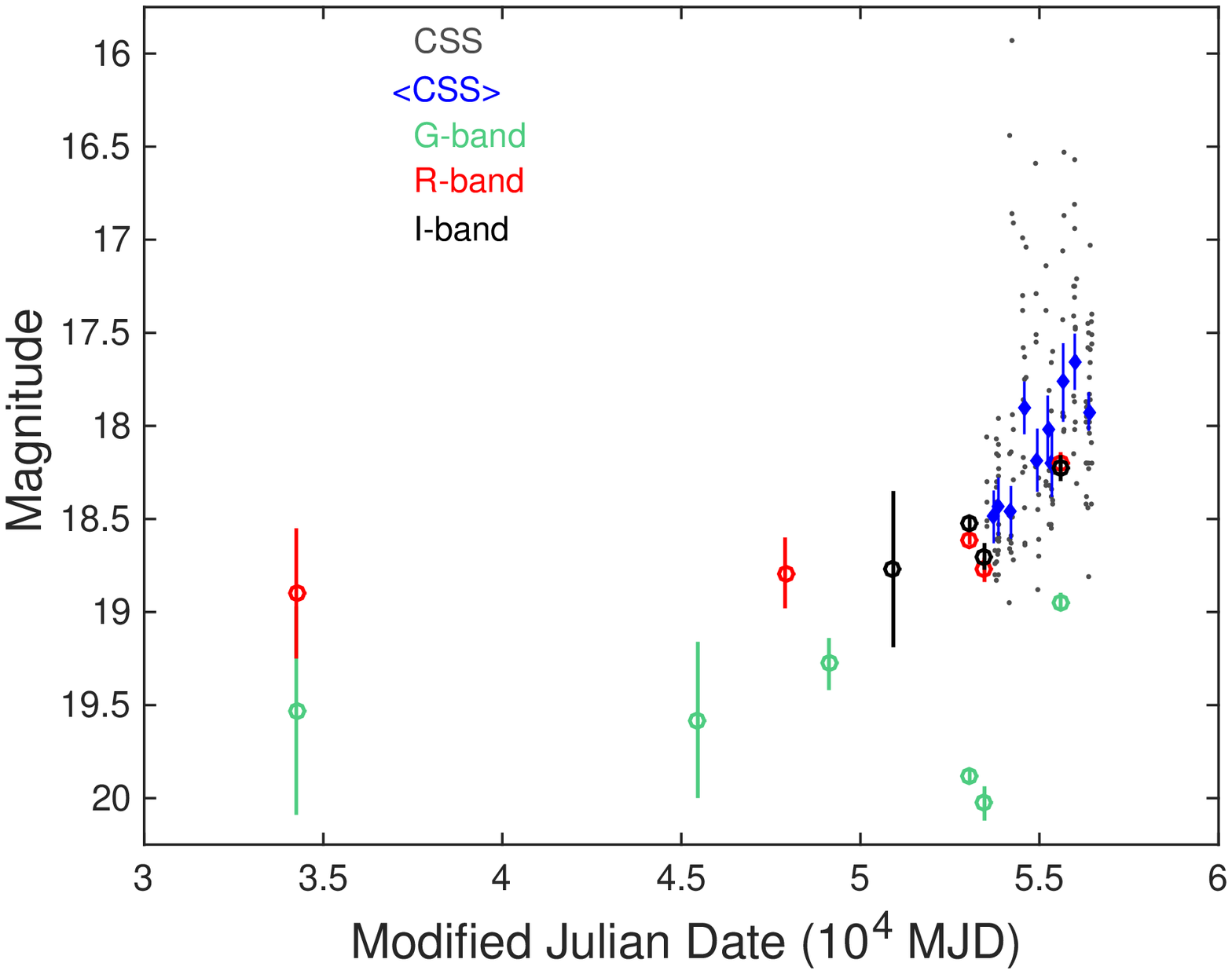}
\caption{\label{fig:f_j12magvsmjd} Above is the light curve of quasar J1231.  The Modified Julian Dates (MJDs) are tabulated in Table \ref{tab:tphot2}. The data are grouped in the $g$, $r$, and $i$ bands indicated by the colours on the graph, as well as in the optical band provided by the Catalina Sky Survey, CSS and the calculated averaged CSS, $<$CSS$>$.}
\end{figure}

\begin{table*}
\footnotesize
\caption{Equivalent Width Differences$^a$}
\label{tab:DEW1}
\begin{tabular}{lccccc}
\hline
Quasar	&	J0841			    &	J1231			    &	J1408			&	J1408			&	J1408			\\
\hline																					
Epoch~1	&	53705.97			&	53466.47			&	53795			&	53795			&	...			\\
Epoch~2	&	55565.82			&	55591.38			&	55276			&	...			    &	55276			\\
Epoch~3	&	...			        &	...			        &	...			    &	56339			&	56339			\\
Redshifts~used	&	2.337			&	1.0227			&	0.834			&	0.834			&	0.834			\\
$\Delta EW_{l}$(\mgii)$^b$	&	-29.02	$\pm$	1.00	&	-14.38	$\pm$	1.54	&	-17.24	$\pm$	0.39	&	-31.28	$\pm$	0.36	&	-14.04	$\pm$	0.50	\\
$\Delta EW_{l}$(\feii)$^b$	&	-42.59	$\pm$	0.83	&	-33.53	$\pm$	1.50	&	-18.92	$\pm$	0.25	&	-26.45	$\pm$	0.24	&	-7.53	$\pm$	0.32	\\
$\Delta EW_{l}$(\mgii+\feii)$^b$	&	-64.08	$\pm$	1.18	&	-38.92	$\pm$	1.74	&	-36.54	$\pm$	0.45	&	-57.84	$\pm$	0.42	&	-21.30	$\pm$	0.58	\\
\hline		
\end{tabular}
\normalsize
\\
$^a$ All equivalent widths are measured in \AA.\\
$^b$ Measured with respect to a line connecting the two ends of the trough in the later epoch \citep{fbqs1408}.\\
\end{table*}

\begin{table*}
\footnotesize
\caption{Photometry of SDSS J123103.70+392903.6}
\label{tab:tphot2}
\begin{tabular}{|cccccccc|}
\hline
Date       & MJD      & Source  & Original         & Syn.           & $g_{est}$        & $r_{est}$        & $i_{est}$     \\ \hline 
1952-08-22 & 34246.23 & POSS-I  & O=20.18$\pm$0.52 & 20.53$\pm$0.17 &  19.53$\pm$0.55  & ...              & ...           \\ 
1952-08-22 & 34246.23 & POSS-I  & E=18.59$\pm$0.30 & 18.30$\pm$0.15 & ...              &  18.90$\pm$0.34  & ...           \\
1983-05-07 & 45461.24 & Palomar & V=18.80$\pm$0.41 & 19.19$\pm$0.05 &  19.58$\pm$0.41  & ...              & ...           \\
1993-05-13 & 49120.19 & POSS-II & J=18.50$\pm$0.10 & 19.10$\pm$0.09 &  19.28$\pm$0.13  & ...              & ...           \\
1990-01-05 & 47896.49 & POSS-II & F=18.73$\pm$0.14 & 18.55$\pm$0.12 & ...              &  18.79$\pm$0.18  & ...            \\ 
1998-04-15 & 50918.33 & POSS-II & N=19.07$\pm$0.38 & 18.82$\pm$0.16 & ...              & ...              &  18.77$\pm$0.41  \\
2004-02-17 & 53052.44 & SDSS    & $g,r,i$ PSF      & ...            & 19.876$\pm$0.024 & 18.613$\pm$0.015 & 18.521$\pm$0.026 \\
2005-04-06 & 53466.47 & SDSS    & spectrum         & ...            & 20.029$\pm$0.082 & 18.774$\pm$0.055 & 18.701$\pm$0.062 \\
2011-01-30 & 55591.38 & BOSS    & spectrum         & ...            & 18.945$\pm$0.038 & 18.204$\pm$0.052 & 18.227$\pm$0.060 \\
\hline
\end{tabular}
\end{table*}
\normalsize

Referring to Figure \ref{fig:f_j12magvsmjd}, the $g$, $r$, and $i$ data was taken from POSS-I, POSS-II, Palomar, SDSS, and BOSS as indicated in Table \ref{tab:tphot2} and as described in \S~\ref{histphot} above. This quasar is much dimmer in magnitude compared to J0841, however the trends in the data are very similar. The $g$, $r$, and $i$ bands all follow similar patterns as Figure \ref{fig:f_j08magvsmjd}. The CRTS data (see \S\ref{Phot}) is represented by the grey points and its averaged values are indicated in the blue. Since there are many evident outliers in this data, magnitudes brighter than 17 were neglected in the average value calculations. There are 10 bins each with a size of approximately 16 data points in length. The trend is linear however it still follows a similar projection as the CRTS data in Figure \ref{fig:f_j08magvsmjd}.


\subsection{Quasar J1408} \label{j1408}

The dramatic absorption variability in a time span less than 5 rest-frame years of this quasar was first discussed in \cite{fbqs1408}.

\begin{figure} 
\centering
\includegraphics[width=\columnwidth]{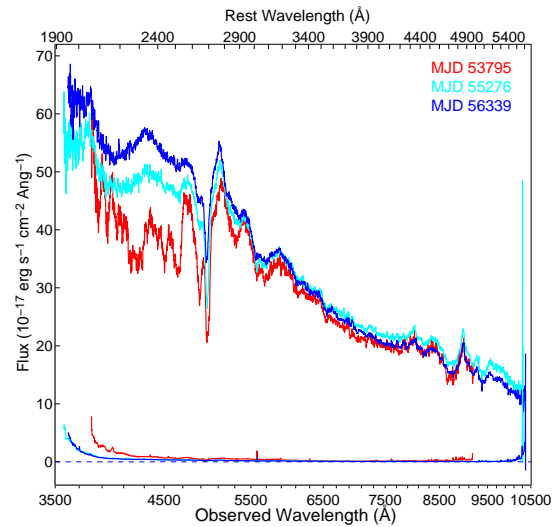}
\caption{\label{fig:f_j1408}Spectra of FBQS J1408 showing continued decrease in absorption across three epochs. The corresponding curves on the bottom are the 1-$\sigma$ uncertainties of the flux for all three epochs. The bottom axis shows observed-frame wavelengths, and the top axis shows rest-frame wavelengths. The top axis is labeled using a redshift of $z = 0.848$. The tick marks of the top axis are every 100 \AA\ with labels every 500 \AA. The tick marks of the bottom axis (facing out) are every 100 \AA\ with labels every 1000 \AA. The spectra are smoothed with a 5 pixel box-car weighted average.}
\end{figure}

We present the continued decrease of this object's absorption in Figure \ref{fig:f_j1408}.
As the absorption has conntiued to decrease, the underlying power-law spectrum of the quasar is quite obvious in these latest epochs.

\subsubsection{Power Law Fitting and Analysis} \label{j1408plaw}

No spectrophotometric correction is needed for the BOSS spectra of this object.
To analyze absorption and continuum changes, for each spectrum we find the best-fit amplitude and slope of a power law continuum 
$F_\lambda=A(\lambda/3400.7$~\AA$)^\alpha$,
with $\lambda$ being the rest-frame wavelength in \AA.
The fitting regions used were $3021 \leq \lambda \leq 3100$~\AA, $4145 \leq \lambda \leq 4157$~\AA, and $4202 \leq \lambda \leq 4227$~\AA ~(the same as J1231).
The index fits in ascending order of our three epoch are $-1.6$, $-1.5$, and $-1.8$. We estimate the systematic uncertainty to be $\pm 0.2$ for each fit. The associated $\chi^{2}_{\nu}$ values are approximately 0.748, 0.678, and 0.561. These power law fits can be used to estimate black hole mass, however we adopt for continuity, the value given by \cite{fbqs1408}.

\subsubsection{Black Hole Mass Estimate} \label{j1408mbh}

In \cite{fbqs1408}, we found that this object has
$M_{BH}=(3.15 \pm 0.35) \times 10^9$~$M_\odot$,
which yields a Schwarzschild radius of
$R_{Sch}=9.3 \times 10^9$\,km.
We also found $L_{bol} = (2.74\pm 0.62) \times 10^{46} {\rm ~erg~s^{-1}}$
which yields $L_{bol}/L_{edd} = 0.07 \pm 0.02$ for this object,
using the mean bolometric correction of \cite{gtr06}.
Note that the accretion disk size for J1408 given in \cite{fbqs1408} is in error.
The diameter $D_{2700}$ within which 90\% of the 2700~\AA\ continuum is emitted for a non-rotating black hole with the parameters given above for J1408 is 68 $R_{Sch}$, or $6.33\times 10^{11} {\rm ~km}$. The black hole mass and Eddington ratio were used to calculate transverse velocities of bulk motion in our discussion.

\subsubsection{Trough Identification} \label{j1408ids}

In the later BOSS spectrum of J1408, the only absorption present is from Mg\,{\sc ii}; Fe\,{\sc ii} is seen in emission, not absorption.  We also confirm the finding of \cite{2015ApJ...803...58Z} that H$\beta$ absorption was present in J1408 up to and including the SDSS epoch.

\subsubsection{Equivalent Width}
\label{EW1408}
The equivalent width of \feii\ absorption decreases for about $18.92\pm 0.25$~\AA\ from $2006$ to $2010$ while \mgii\ equivalent width decreases at the same time for about $17.24\pm 0.39$~\AA. From $2010$ to the last epoch in $2013$ (BOSS spectrum), the \feii\ absorption decreases slowly for about $7.53\pm 0.32$~\AA\ while \mgii\ equivalent width decreases dramaticaly for about $14.04\pm 0.50$~\AA, by a factor of $3$ faster than \feii\ decrease in two earlier epoachs, see column 4 and 6 in Table \ref{tab:DEW1}.

\subsubsection{Historical Photometry} \label{j1408phot}

\begin{figure} 
\centering
\includegraphics[width=\columnwidth]{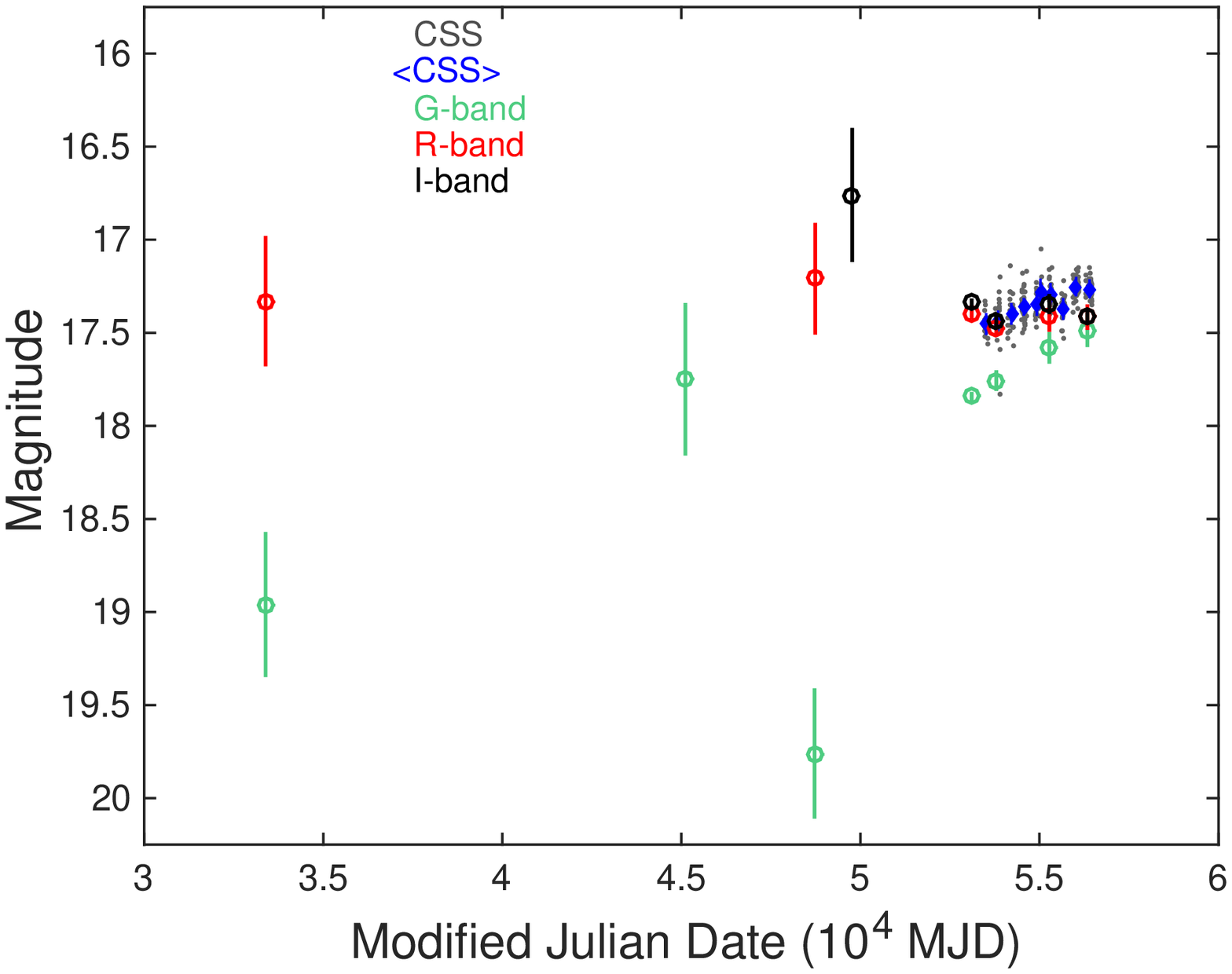}
\caption{\label{fig:1408magvsmjd}Above is the light curve of quasar J1408.  The Modified Julian Dates (MJDs) are tabulated in Table \ref{tab:tphot3}. The data are grouped in the $g$, $r$, and $i$ bands indicated by the colours on the graph, as well as in the optical band provided by the Catalina Sky Survey, CSS and the calculated averaged CSS, $<$CSS$>$.}
\end{figure}

We provide updated synthetic photometry for J1408 in Table \ref{tab:tphot3}.
From comparison of estimated $g$ magnitudes to those measured in 2004 (the SDSS imaging epoch) when the absorption was weakening,
we infer that the absorption was stronger in the photographic epochs, with the possible exception of the Palomar Quick V epoch (see Fig. \ref{fig:1408magvsmjd}).
The estimated $g$ magnitude in that epoch is consistent with the value in the SDSS epoch and is brighter than found in the POSS-I and POSS-II epochs
at 2.2$\sigma$ and 3.8$\sigma$ significance, respectively.
Unfortunately, with only one discrepant magnitude we cannot confirm that it represents a period of decreased absorption in the early 1980s.
(We did not identify the Quick V epoch as discrepant in \cite{fbqs1408} because in that work we assumed a red $B-V$ colour and used our highly uncertain $V$ magnitude to estimate the $g$ magnitude at the quick $V$ epoch.
Here, in contrast, we use the accurate SDSS $g$ and $r$ magnitudes to synthesize the $V$ magnitude that would have been measured in the SDSS epoch.)

\begin{table*}
\footnotesize
\caption{Photometry of FBQS J140806.2+305448}
\label{tab:tphot3}
\begin{tabular}{|cccccccc|}
\hline
  Date       & MJD      & Source  & Original         & Syn.           & $g_{est}$        & $r_{est}$        & $i_{est}$     \\ \hline 
  1950-04-18 & 33389    & POSS-I  & O=19.24$\pm$0.34 & 18.12$\pm$0.17 &  18.96$\pm$0.38  & ...              & ...            \\
  1950-04-18 & 33389    & POSS-I  & E=17.10$\pm$0.30 & 17.17$\pm$0.15 & ...              &  17.33$\pm$0.34  & ...            \\
  1982-05-21 & 45110    & Palomar &  V=17.5$\pm$0.4  & 17.59$\pm$0.05 &  17.75$\pm$0.40  & ...              & ...            \\
  1992-04-08 & 48720    & POSS-II & J=19.31$\pm$0.33 & 17.39$\pm$0.09 &  19.76$\pm$0.34  & ...              & ...            \\
  1992-04-26 & 48738    & POSS-II & F=17.26$\pm$0.26 & 17.45$\pm$0.12 & ...              &  17.21$\pm$0.29  & ...            \\
  1995-02-23 & 49771    & POSS-II & N=17.06$\pm$0.31 & 17.64$\pm$0.16 & ...              & ...              &  16.76$\pm$0.35 \\
  2004-04-13 & 53108    & SDSS    & $g,r,i$ PSF      & ...            & 17.843$\pm$0.015 & 17.404$\pm$0.013 & 17.339$\pm$0.013 \\
  2006-03-01 & 53795    & SDSSsp  & ...              & ...            & 17.757$\pm$0.046 & 17.474$\pm$0.037 & 17.442$\pm$0.029 \\
  2010-03-21 & 55276    & BOSSsp  & ...              & ...            & 17.581$\pm$0.075 & 17.407$\pm$0.078 & 17.346$\pm$0.044 \\
  2013-02-16 & 56339    & BOSSsp  & ...              & ...            & 17.485$\pm$0.082 & 17.416$\pm$0.059 & 17.411$\pm$0.035 \\
\hline
\end{tabular}
\end{table*}
\normalsize

\subsection{SEDs} \label{sed}

\begin{figure} 
\centering
\includegraphics[width=\columnwidth]{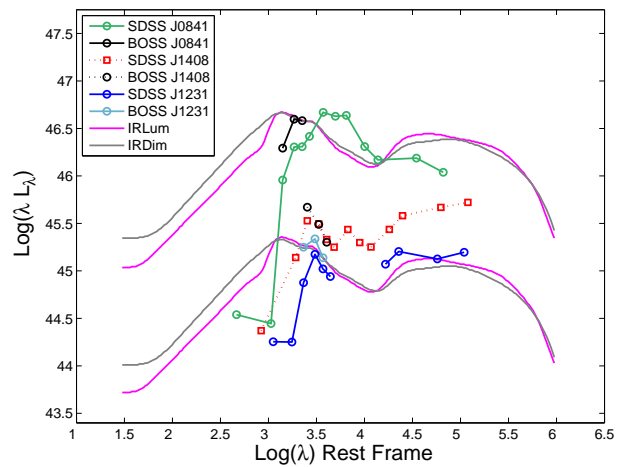}
\caption{\label{fig:SED}
Above is a rest-frame spectral energy distribution (SED) plot of the SDSS and BOSS data for J0841 in green and black, J1231 in the dotted red line and black line, and J1408 in blue and cyan. The infrared luminous and infrared dim quasar SEDs from Richards et al.\ (2006) are also plotted in pink and black respectively, positioned with respect to the BOSS $i$-band for J0841 and J1231. All of the BOSS data points are shown in the $g$, $r$, $i$ bands.}
\end{figure}

Figure \ref{fig:SED} displays the SEDs in the rest frame for J0841, J1231, and J1408 using their respective redshifts from \S~\ref{redshift}. The data points are the UV range from GALEX (J0841 and J1408 are in the fUV and J1231 was calculated in the nUV) followed by $u,g,r,i,z$ from SDSS, then J, H, K from 2MASS and the final four points from WISE. J0841 and J1408 both have points in the J, H, K bands from 2MASS, however J1231 does not contain data in those bands hence the gap in its plot. Refer to Table \ref{tab:tphot4} for further detail on this data.

The infrared-luminous and infrared-dim SEDs shown were constructed in \cite{gtr06} by dividing their sample at the median luminosity in the rest-frame $1-100~{\mu \rm m}$ band.
These data sets allow comparisons of each quasar to be made regarding their brightness in the infrared. The final four infrared points are the key aspects to focus on.

As J0841 clearly falls beneath the magenta and grey, it suggests the quasar is underluminous in the infrared. The slope of the three shortest-wavelength WISE data points for J0841 is $-1.44$. This is a very unusual slope in comparison to the slopes of $>-0.7$ found for essentially all the 2,964 $z\simeq 2$ BAL quasars of \cite{2014ApJ...786...42Z} in their Fig.\ 2.
J1231 differs in this regard as it lies slightly above the magenta and grey lines, which indicates it is fairly infrared bright but not intensely so. If one were to shift the magenta and grey plots up to match the J1408 $i$-band point, it would appear this quasar is more infrared luminous than J1231.

The SEDs of the three quasars do not display similar trends and thus no explicit correlation between the SEDs and the outflows can be made.
Variability in the optical rest-frame, due to the different epochs at which the datasets were obtained, affects the relative position of the optical with respect to the infrared and consequently the normalization of the SEDs plotted for comparison.
However, the BOSS and WISE data points were taken at similar epochs, and so optical variability is unlikely to explain why in the infrared one of our quasars (J0841) appears underluminous and the other two appear more luminous than average.


\section{Discussion} \label{discuss}

\subsection{Implications} \label{imply}

We have reported three cases of FeLoBAL quasars whose Fe\,{\sc ii} absorption has weakened dramatically.
\cite{mcgraw+2015} report the same for SDSS J120337.92+153006.7.
\cite{2015ApJ...803...58Z} report the same for FBQS J072831.6+402615 and also report that FBQS J152350.4+391405 is an object wherein the Fe\,{\sc ii} and Mg\,{\sc ii} absorption strengthened between two SDSS epochs and has stayed strong since (their Fig.\ 2).

The detection to date of five overlapping-trough FeLoBAL quasars whose Fe\,{\sc ii} absorption has weakened dramatically and only one where it has strengthened raises the question:
does the sample from which they were drawn contain an equal number of FeLoBALs in which Fe\,{\sc ii} absorption has strengthened dramatically?

The answer may depend on the cause of the variability.
If we are observing fluctuations in winds due to motion of absorbing structures across our line of sight (or due to short-term variable ionizing flux with no consistent long-term trend), then equal numbers of quasars with weakening and strengthening Fe\,{\sc ii} absorption are expected.
(The possible past epoch of decreased absorption in J1408 discussed in \S~\ref{j1408phot} is possible evidence of such fluctuations.)

If we are seeing winds that are transitioning permanently to higher ionization, then whether or not we see quasars with strengthening \feii\ absorption depends on how strong the typical reddening from dust is in such quasars.
If quasars with strengthening \feii\ absorption represent the birth of FeLoBAL outflows, and if such outflows are all born in heavily dust reddened quasars, then such quasars will be underrepresented in optical quasar surveys.
As a result, optical quasar surveys will show an excess of FeLoBAL quasars with weakening \feii\ absorption.  For example, Figure 1 of \cite{mcgraw+2015} shows twelve FeLoBAL quasars whose normalized spectra show strengthening absorption in only one region of one spectrum (SDSS J142703.60+270940.0), although their Figure A does show that SDSS J121442.30+280329.0 showed both slightly weakening and slightly strengthening absorption when observed at multiple epochs.

\subsection{Transverse Velocities} \label{vtrans}

Assuming the vanishing \feii\ in these objects is due to the motion of the BAL outflow across our line of sight, some constraints may be placed on the transverse velocity of the BAL outflow.

Our single rectangular flow tube (SRFT) simulations consist of the following. First, a square space the width of the accretion disk is created and populated with flux data at a specific wavelength.
This is done assuming that the temperature of the disk is given by Eq.~4.23 of \cite{net13} and the disk is a black-body radiator.
Then, horizontal clouds are swept across the square space assuming full opacity to flux. The horizontal clouds vary in width and
closest approach to the black hole.
The maximum and minimum distance a cloud can travel to go from one covering fraction (CF) to another seen in the SDSS epochs gives us a maximum and minimum velocity of the clouds.
The CF was measured by plotting the best fit power-laws for each spectrum and calculating the fraction of flux missing at the specific wavelength according to the power-law fit. We round to 1 significant digit to represent the significant uncertainty.

In the case of J0841, the smooth `S curve' change in magnitude seen in CRTS photometry corresponds to simulations of SRFT across the accretion disk.
Using simulations of SRFT, the change in flux of the \feii\ absorption pattern around 2525~\AA~rest-frame seen in the two epochs (going from a covering factor of $\simeq$0.6 to $\simeq$0.2 in 556 rest-frame days) could be caused by transverse motion at velocities roughly $10,000$ \kms\ to $42,000$ \kms.
Using the simple outflow model of \cite{fbqs1408}, we obtain a very similar range of $11,000$ \kms\ to $34,000$ \kms.
We also obtain a constraint of $4700 ~R_{Sch} < d_{BAL} < 42,000 ~R_{Sch}$ for the distance of the BAL region in J1231 from the continuum source.

Because the CRTS photometry for J1231 has more scatter, and because J1231 showed increased continuum flux as well as weakening BAL troughs, there is more uncertainty about a transverse motion explanation for its behavior than for that of J0841.  Nonetheless, transverse motion is not ruled out, so we quote the results of applying the same models for that scenario to J1231.
We find that the change in flux of \feii\ absorption around 2500~\AA~rest-frame (going from a covering factor of $\simeq$0.7 to 0.3 in 1146 rest-frame days) could be caused by transverse motion at velocities roughly $1000$ \kms\ to $4000$ \kms. (Our inferred velocities are larger than those of \cite{mcgraw+2015} for J1231 because we adopt a larger $M_{BH}$ and because we do not assume a uniformly emitting continuum source.)
Using the simple outflow model of \cite{fbqs1408}, we obtain a very similar range of $1000$ \kms\ to $5000$ \kms.
We also obtain a constraint of $10^4 ~R_{Sch} < d_{BAL} < 8\times 10^4 ~R_{Sch}$ for the distance of the BAL region in J1231 from the continuum source.

For J1408, using the corrected disk size of \S~\ref{j1408mbh} and the simple outflow model of \cite{fbqs1408}, we obtain a transverse velocity between 4,400 \kms\ and 32,400 \kms.  (The large range arises because the timescale over which the absorption varied is only constrained to between $226<t<1849$ rest-frame days.)
We also obtain a constraint of $5400 ~R_{Sch} < d_{BAL} < 42,000 ~R_{Sch}$ for the distance of the BAL region in J1231 from the continuum source.

Note that all the transverse velocities given here are lower limits in the sense that microlensing observations indicate that quasar accretion disks are a factor of $\sim$4 larger in diameter than predicted by the standard accretion disk theory used in our calculations \citep{2010ApJ...712.1129M}.
Higher transverse velocities require the BAL absorbers be located at smaller distances from the central black hole.
Relatively small distances (and relatively high densities) for the absorbers in overlapping-trough FeLoBAL quasars are also indicated by the photoionization study of \cite{2015ApJ...803...58Z}, but see also \cite{2014ApJ...783...58L}.

\subsection{Ionization Variability} \label{ionvar}

Variability in the ionization state of the absorbing gas was considered as an explanation for dramatic \feii\ variability in \cite{fbqs1408}, but was deemed unlikely.  Since that time, however, it has been shown that at least half of \civ\ BAL variability is coordinated between widely separated troughs in the same object; see \S~4.7 of \cite{FB13}.
Such variability is most naturally explained by ionization variability in the absorbing gas.

The level of near-UV continuum variability of a quasar unfortunately does not constrain its level of ionizing flux variability, because quasars are more variable in flux at shorter wavelengths; see the discussion and references in \S~4.2.1 of \cite{2015ApJ...806..111G}.
Thus, even though only one among the three quasars in this paper shows an increase in near-UV continuum flux accompanying the disappearance of \feii\ absorption, ionization variability remains a plausible explanation for the absorption changes.

There is circumstantial evidence for photoionization variability in J1231, as both the continuum and the absorption have changed. For example, an increase in extreme-ultraviolet ionizing flux accompanying the observed near-ultraviolet flux increase may ionize enough \feii\ to \feiii\ to change the \feii\  absorption features. However, due to the many unconstrained parameters for such a scenario in this object, modelling these changes is beyond the scope of this paper.

\subsection{Mg II and UV Fe II Emission Line Blueshifts} \label{bshifts}

\begin{figure*}
\centering
\includegraphics[width=13cm,height=5cm]{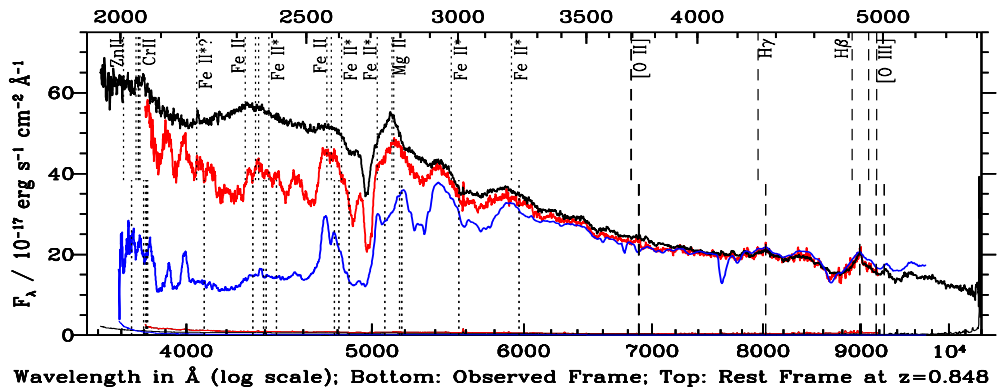}
\caption{\label{j1408plog}
The blue and red curves are the Keck and SDSS spectra of J1408, respectively, from \citet{fbqs1408} and the black curve is the later of the two BOSS spectra. All spectra are smoothed by a 5-pixel boxcar, with 1-$\sigma$ uncertainties shown along the bottom. The top axis shows the rest frame at $z=0.848$ (the Balmer-line and adopted systemic redshift), and vertical lines in the bottom half of the figure show the expected positions of various transitions at that redshift. Vertical lines in the top half of the figure show the expected positions of those same transitions at $z=0.831$, the adopted \mgii\ and UV \feii\ redshift. The significant redshift difference between those lines and the Balmer lines is apparent.
}
\end{figure*}

As mentioned in passing in \S~2.1 of \cite{fbqs1408}, the Mg\,{\sc ii} emission peak in J1408 is blueshifted from the Balmer-line systemic redshift of $z=0.848\pm 0.001$.
BOSS spectra of J1408 also confirm blueshifts for UV Fe\,{\sc ii} emission lines shortward of $\sim$3000~\AA, which arise from energy levels $\leq 1$~eV above the Fe\,{\sc ii} ground state, but not for longer-wavelength Fe\,{\sc ii} emission lines which arise from higher-excitation levels.
If we conservatively adopt the DR12Q redshift of $z=0.8310\pm 0.0004$ to represent the Mg\,{\sc ii} redshift, as shown in Figure \ref{j1408plog}, we find a blueshift of $2800 \pm 200$ km~s$^{-1}$.\footnote{A blueshift of up to $3800 \pm 300$ km~s$^{-1}$ is possible if we adopt the peak Mg\,{\sc ii} redshift of $z=0.825\pm 0.002$.}
This is an extremely large blueshift for Mg\,{\sc ii}, as can be seen from Fig.\ 13 of \citep{bossdr9q}.
This blueshift for Fe\,{\sc ii} contradicts the tendency in low-redshift quasars for Fe\,{\sc ii} to be redshifted with respect to H$\beta$ \citep{2008ApJ...687...78H}.

To determine if J1408 is exceptional in this regard, or if other overlapping-trough BAL quasars have similar velocity offsets which are masked by very strong absorption cutting into their Mg\,{\sc ii} emission lines, we inspected four other overlapping-trough FeLoBAL quasars with coverage of H$\beta$ in SDSS or BOSS spectroscopy.  In three cases\footnote{SDSS J090212.42+592415.8, J151130.30+140551.0, and J153420.23+413007.6.} the Mg\,{\sc ii} emission was too strongly affected by absorption to determine its velocity offset.

\begin{figure}
\centering
\includegraphics[width=\columnwidth]{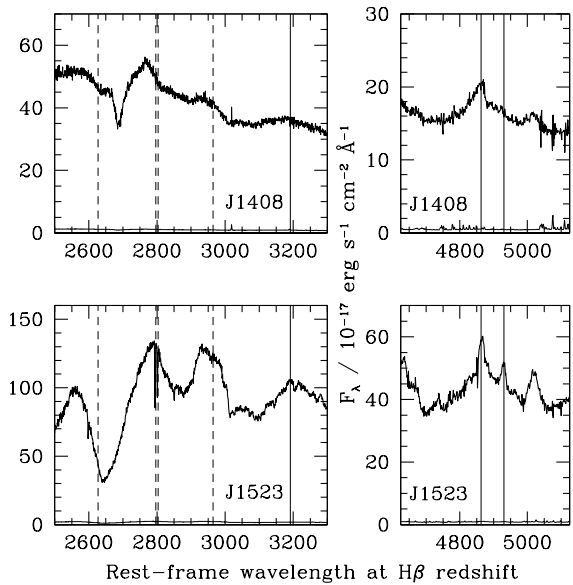}
\caption{\label{fig:bs}
Blueshifted \mgii\ and UV \feii\ emission.
The top panels show J1408, with vertical lines at the expected rest wavelengths at $z=0.848$.
Similarly, the bottom panels show J1523 at $z=0.6609$.
The left panels show the regions around \mgii\ and the right panels the regions around H$\beta$,
all at the same wavelength scale.
In each quasar, H$\beta$ and \feii\ emission lines at $\geq$3000~\AA\ (solid lines)
are found at the systemic redshift.
In contrast, \mgii\ and \feii\ emission lines at $\leq$3000~\AA\ (dashed lines)
are clearly blueshifted in J1408, and the \mgii\ line is tentatively blueshifted in J1523.
The \feii\ emission line wavelengths are taken from Vanden Berk et al.\ (2001).
}
\end{figure}

For the fourth case, SDSS J152350.4+391405 (J1523), we measure a tentative Mg\,{\sc ii} blueshift of $1100 \pm 300$ km~s$^{-1}$ relative to the H$\beta$ redshift of $z=0.6609\pm 0.0022$.
The emission-line blueshifts of this object and J1408 are shown in Figure \ref{fig:bs}.
The blueshift in J1523 is tentative because the emission in that object is also affected by broad Mg\,{\sc ii} and Fe\,{\sc ii} absorption (Fig.\ 2 of Zhang et al.\ 2015), although any decrease in absorption seems likely to reveal a larger blueshift, as happened in J1408 --- see \cite{fbqs1408} Figure 2 and \S~2.1.
In addition, Zhang et al.\ (2015) used their own follow-up spectroscopy of FBQS J072831.6+402615 to show it is another example of an FeLoBAL quasar with vanishing Fe\,{\sc ii} absorption.
Their Fig.\ 1 appears to show that this quasar has an even larger blueshift of its
Mg\,{\sc ii} emission relative to H$\beta$, by about 6000~km\,s$^{-1}$.

We tentatively conclude that the emission lines of \mgii\ and of UV \feii\ transitions are blueshifted by over a thousand \kms\ relative to the \Hb\ redshift in FeLoBAL quasars with substantially variable \feii\ absorption.
The most straightforward interpretation is that those emission lines arise predominantly in outflowing gas with a substantial velocity component along our line of sight in those objects.
If that interpretation is correct, the continued presence of emitting gas flowing toward us in these quasars would imply that their \feii\ absorption variability is likely due to transverse motion of gas out of our line of sight.  If the absorption variability is due to changing ionization, more lines of sight than just ours would be affected but we would see the effects on those sightlines later (due to geometric time delays), and blueshifted emission lines should weaken over time as we see the wind outflowing in other directions being shifted to higher ionization.  Spectral monitoring of these quasars should be able to distinguish between these possibilities.

It is not known if other FeLoBAL quasars or non-BAL quasars with strong \feii\ emission exhibit a similar \mgii\ and UV \feii\ emission-line blueshift.  Any such shift in non-BAL quasars will be typically lower in magnitude if non-BAL lines of sight are typically found at larger angles to the outflow directions.

\section{Conclusions} \label{concl}

Our conclusions are the following.

$\bullet$~ We report a dramatic decrease in iron absorption strength in the iron low-ionization broad absorption line quasar SDSS J084133.15$+$200525.8.

$\bullet$~ We report on the continued weakening of absorption in the prototype of this class of variable broad absorption line quasar, FBQS J140806.2$+$305448.

$\bullet$~ We also report a third
example of this class, SDSS J123103.70$+$392903.6; unlike the other two examples,
it has undergone an increase in observed continuum brightness (at 3000~\AA\ rest-frame) as well as a decrease in iron absorption strength.

$\bullet$~ We note that the \mgii\ and UV \feii\ lines in several FeLoBAL quasars are blueshifted by thousands of \kms\ relative to the H$\beta$ emission line peak. We suggest that such emission arises in the outflowing winds normally seen only in absorption.

$\bullet$~ We conclude that these changes could be caused by absorber transverse motion or ionization variability.


$\bullet$~ Finally, to help distinguish between the two scenarios (ionization variability and transverse motion of the BAL outflow), we suggest two systematic concurrent spectroscopic observations in the X-ray and visual bands before and after the disappearance of \feii\ absorption lines in potential candidates like J1231, J1408 and J0481.
For example, such observations could be triggered if an \feii\ vanishing episode was caught early enough in its evolution.
Spectroscopic monitoring of quasars with known \feii\ vanishing episodes may also help to distinguish between these possibilities.



\section*{Acknowledgements}

PBH is supported by NSERC.  We thank T.\ Rudyk for assembling the historical photometry of J1231, G.\ Richards for drawing the BOSS spectrum of J1231 to our attention, M.\ Gregg for use of the Keck spectrum of J1408, and R.\ Bloch and J.\ Zomederis for assistance in the early stages of this study.

Funding for SDSS-III has been provided by the Alfred P. Sloan Foundation, the Participating Institutions, the National Science Foundation, and the U.S. Department of Energy Office of Science. The SDSS-III web site is http://www.sdss3.org/.

SDSS-III is managed by the Astrophysical Research Consortium for the Participating Institutions of the SDSS-III Collaboration including the University of Arizona, the Brazilian Participation Group, Brookhaven National Laboratory, Carnegie Mellon University, University of Florida, the French Participation Group, the German Participation Group, Harvard University, the Instituto de Astrofisica de Canarias, the Michigan State/Notre Dame/JINA Participation Group, Johns Hopkins University, Lawrence Berkeley National Laboratory, Max Planck Institute for Astrophysics, Max Planck Institute for Extraterrestrial Physics, New Mexico State University, New York University, Ohio State University, Pennsylvania State University, University of Portsmouth, Princeton University, the Spanish Participation Group, University of Tokyo, University of Utah, Vanderbilt University, University of Virginia, University of Washington, and Yale University.

\appendix

\section{Spectral Energy Distributions (SEDs)} \label{seds}

\subsection{Flux vs.\ Wavelength}

\begin{figure} 
\centering
\includegraphics[width=\columnwidth]{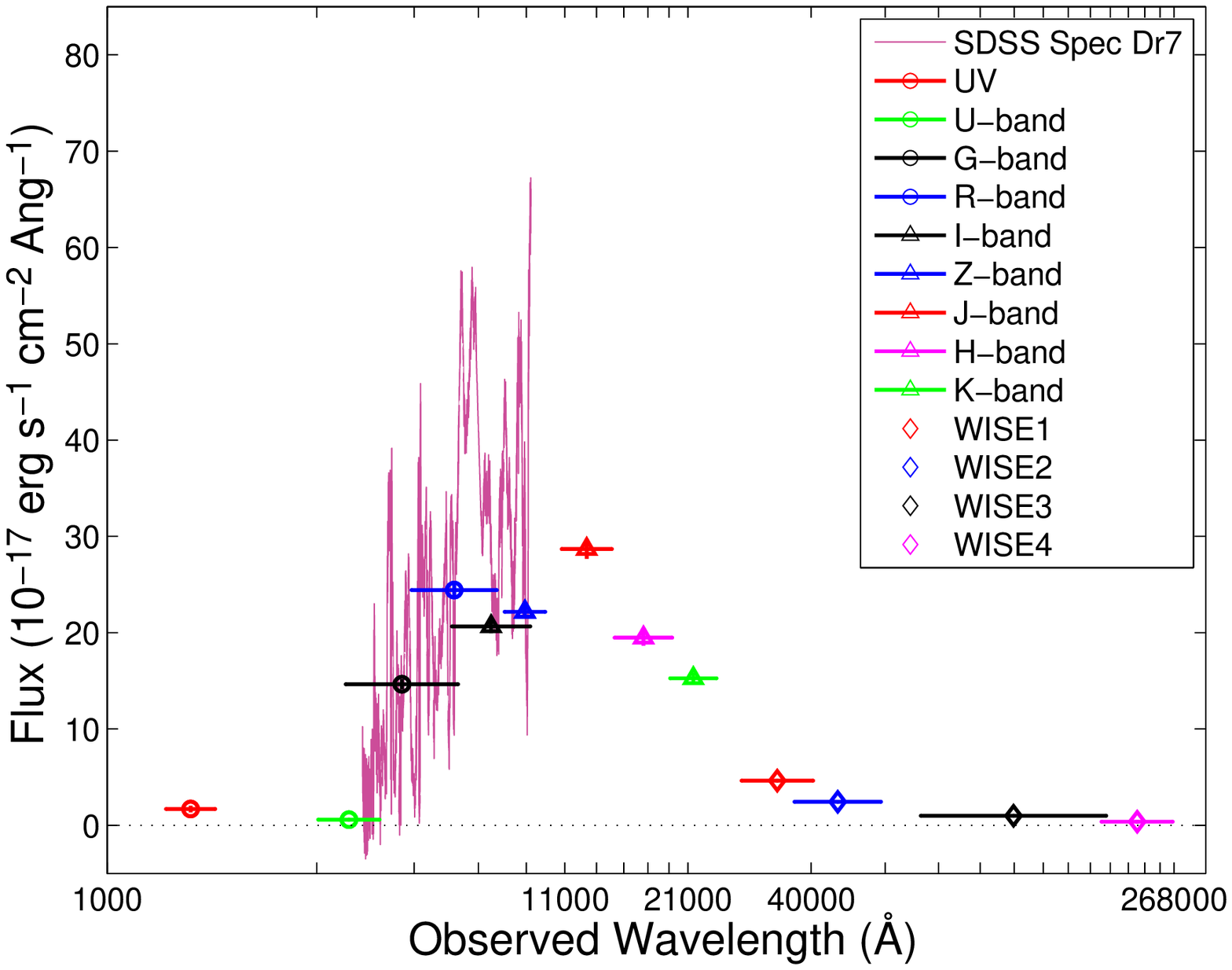}
\caption{\label{fig:f_j08fluxwave}This graph demonstrates the relationship between the observed wavelength and flux of Quasar J0841. The optical spectrum shown in purple was obtained from the SDSS; also shown is the fUV, u, g, r, i, z, j, h, k bands, as well as WISE1, WISE2, WISE3 and WISE4 shown near the trail end of the graph.}
\end{figure}

\begin{figure} 
\centering
\includegraphics[width=\columnwidth]{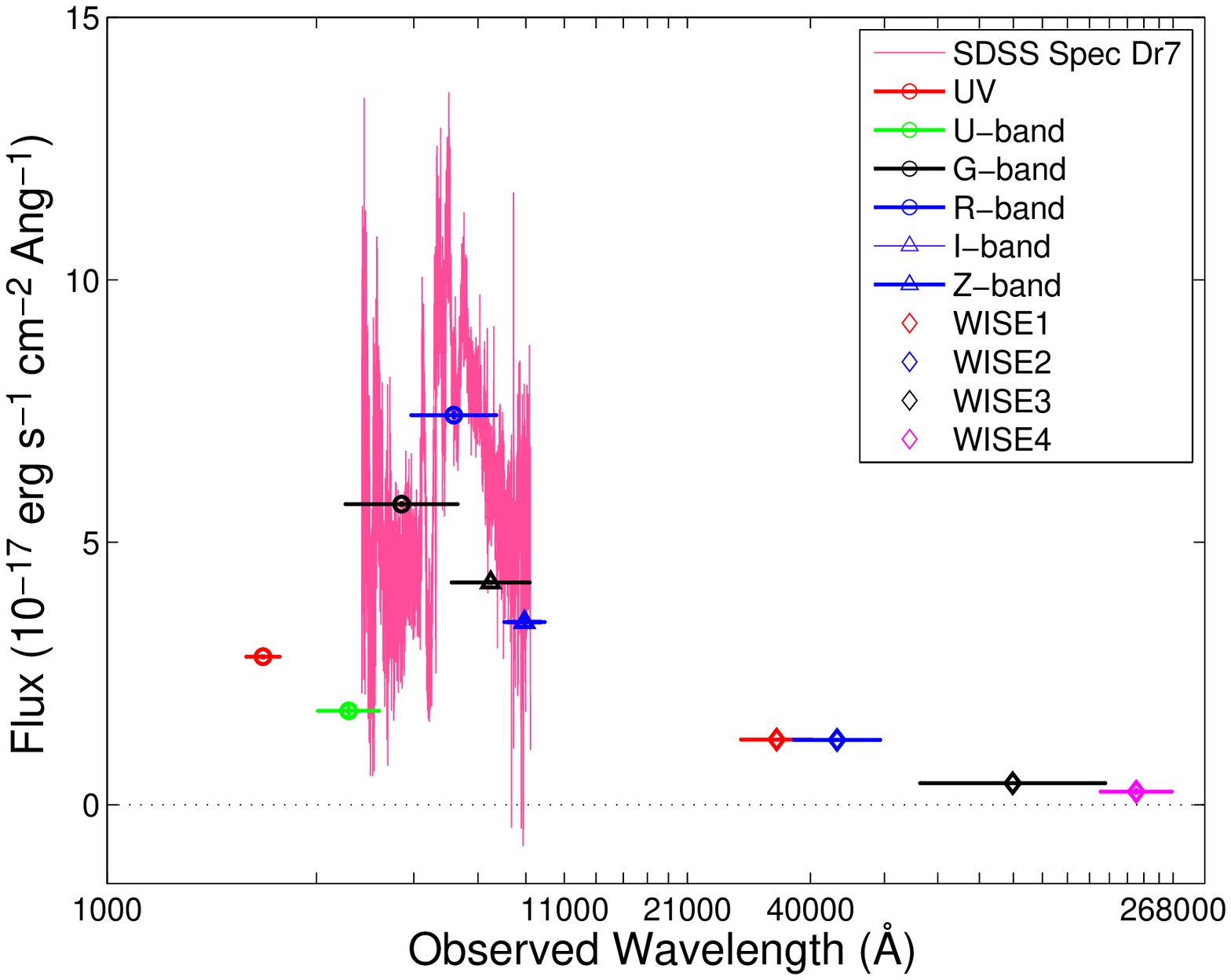}
\caption{\label{fig:f_j12fluxwave}
This graph demonstrates the relationship between the observed wavelength and flux of Quasar J1231. The data are the optical spectrum (SDSS DR7), the nUV, u, g, r, i, and z bands, and the WISE1, WISE2, WISE3, and WISE4 bands shown as the last four points on the graph.}

\end{figure}

\begin{figure} 
\centering
\includegraphics[width=\columnwidth]{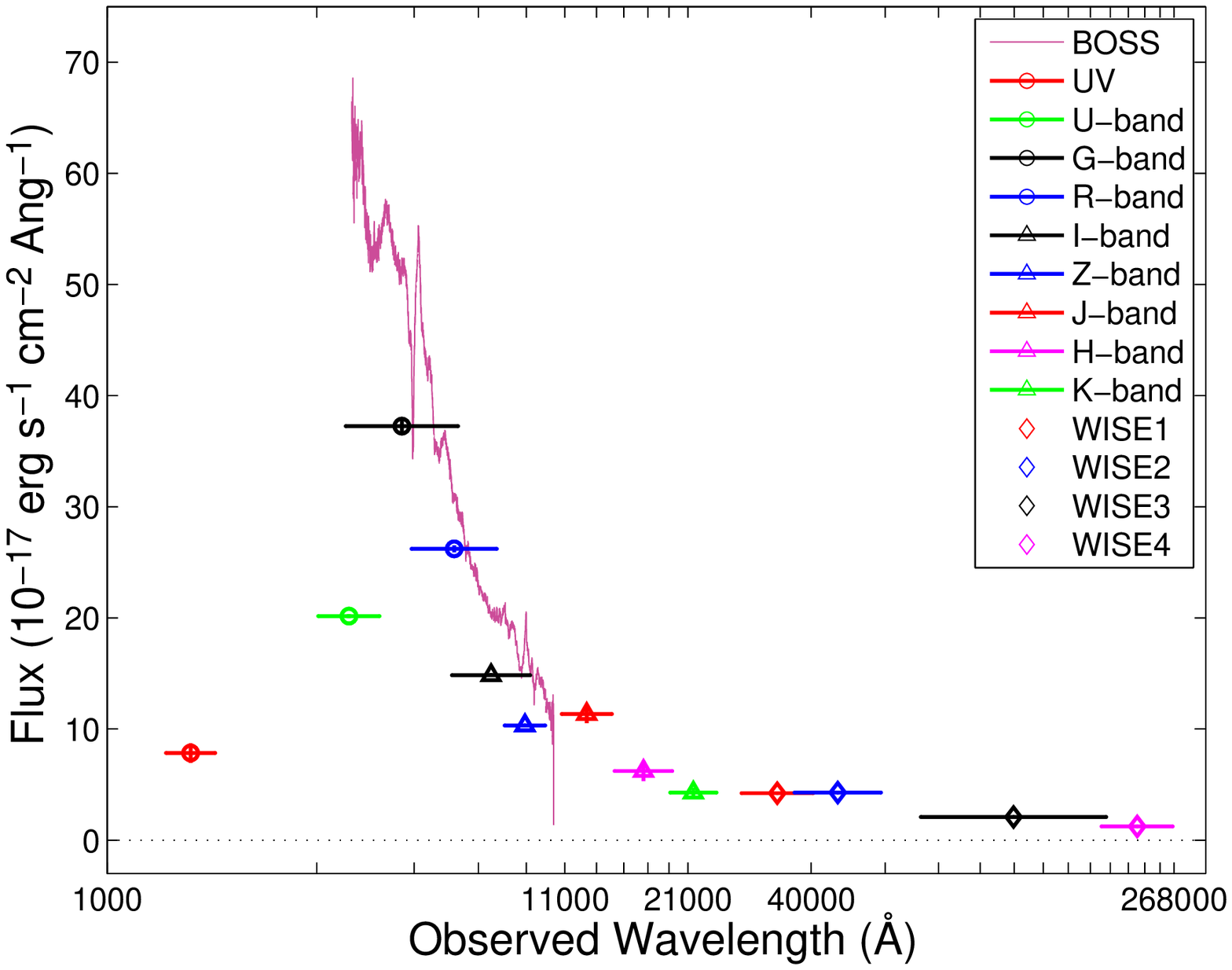}
\caption{\label{fig:f_j14fluxwave}
This graph demonstrates the relationship between the observed wavelength and flux of Quasar J1408. The purple spectrum was retrieved from BOSS, photometric data in the fUV, u, g, r, i, z, j, h, and k bands are displayed, as well as in the WISE1, WISE2, WISE3, and WISE4 bands shown as the last four points on the graph.}

\end{figure}

In order to calculate the flux in Figures \ref{fig:f_j08fluxwave} and \ref{fig:f_j12fluxwave}, we found the designated zero flux and absolute magnitudes (along with their associated uncertainties) from the corresponding sources listed for each filter and telescope system in Table \ref{tab:tphot4}. After that, simple flux-magnitude computations were done to calculate the flux in each of the required bands. The Observed Wavelengths were calculated as a log scale due to the large gap between the wavelengths, i.e starting at 2000 \AA $\:$ in the UV range and ending at 220883 \AA $\:$ in the infrared. Due to this large scaling of wavelengths, the ticks shown for the Observed Wavelengths are not evenly marked, rather they are displayed based on the log scale. It should be noted that \ref{fig:f_j08fluxwave} displays the UV data point in the far-ultraviolet (fUV) range and \ref{fig:f_j12fluxwave} has the UV point in the near-ultraviolet range (nUV). Table \ref{tab:tphot4}  displays the data of the bands with the corresponding wavelengths, zero flux, magnitudes, flux, and data sources for J0841 and J1231.

These two graphs, as well as the tabular data in Table \ref{tab:tphot4}, again demonstrate that J0841 is much brighter in magnitude and flux compared to J1231, hence the lack of 2MASS data (j, h, k bands) for J1231. However, comparing the data of the two graphs similar trends are present.

Figure 4 from \cite{fbqs1408} also displays a flux vs magnitude graph for J1408; an updated and more comprehensive version of these data is displayed in Figure \ref{fig:f_j14fluxwave} and numerically in Table \ref{tab:tphot3}. Comparing \ref{fig:f_j14fluxwave} with \ref{fig:f_j08fluxwave} and \ref{fig:f_j12fluxwave}, J1408 shows complementary results with J0841 and J1231 as the trends in their flux to wavelength relationships show similar patterns. This suggests the changes occurring with these quasars result from an innate property and not due to outside influences.

For Figures \ref{fig:f_j08fluxwave} and \ref{fig:f_j12fluxwave}: the optical spectrum was obtained from the SDSS taken by the seventh data release DR7.

\begin{table*}
\footnotesize
\caption{Data for J0841, J1231, and J1408 in Various Photometric Bands}
\label{tab:tphot4}

\begin{tabular}{|c|c|p{1.8 cm}|p{1.75 cm}|p{2.5 cm}|p{2.25 cm}|p{2.75 cm}|}
\hline

Quasar  & Band  & Wavelength (in \AA)  & Zero Flux$^1$ & Apparent Magnitude   & Flux$^2$ & Observation Date\\ \hline
\multirow{12}{*}{J0841} & fUV$^3$ & 1550 & - & - & $1.700 \pm 0.350$ & 2007-01-24\\
& u$^4$ & 3551 & 367 & $21.970 \pm 0.170$ & $0.598 \pm 0.094$ & 2004-12-12\\
& g$^4$ & 4686 & 511 & $18.856 \pm 0.020$ & $14.656 \pm 0.270$ & 2004-12-12\\
& r$^4$ & 6166 & 240 & $17.481 \pm 0.015$ & $24.424 \pm 0.337$ & 2004-12-12\\
& i$^4$ & 7480 & 128 & $16.980 \pm 0.024$ & $20.664 \pm 0.457$ & 2004-12-12\\
& z$^4$ & 8932 & 78.3 & $16.37 \pm 0.030$ & $22.170 \pm 0.613$ & 2004-12-12 \\
& J$^5$ & 12350 & 31.29 & $15.094 \pm 0.039$ & $28.695 \pm 1.037$ & 1998-10-19 \\
& H$^5$ & 16600 & 11.33 & $14.411 \pm 0.047$ & $19.491 \pm 0.844$ & 1998-10-19\\
& K$^5$ & 21900 & 42.83 & $13.620 \pm 0.039$ & $15.267 \pm 0.548$ & 1998-10-19 \\
& WISE1$^6$ & 33526 & 0.818 & $13.118 \pm 0.013$ & $4.629 \pm 0.0554$ & 2010-10-29 \\
& WISE2$^6$ & 46028 & 0.242 & $12.487 \pm 0.033$& $2.444 \pm 0.074$  & 2010-10-29\\
& WISE3$^6$ & 115608 & 0.00652 & $9.527 \pm 0.063$ & $1.007 \pm 0.058$ & 2010-10-29 \\
& WISE4$^6$ & 220883 & 0.000509 & $7.829 \pm 0.486$ & $0.376 \pm 0.168$  & 2010-10-29\\ \hline
\multirow{9}{*}{J1231} & nUV$^3$ & 2267 & - & - & $2.820 \pm 0.049$ & 2005-09-17\\
& u$^4$ & 3551 & 367 & $20.780 \pm 0.060$ & $1.789 \pm 0.099$ & 2004-02-17\\
& g$^4$ & 4686 & 511 & $19.876 \pm 0.024$ & $5.728 \pm 0.053$ & 2004-02-17\\
& r$^4$ & 6166 & 240 & $18.774 \pm 0.055$ & $7.424 \pm 0.080$ & 2004-02-17\\
& i$^4$ & 7480 & 128 & $18.701 \pm 0.062$ & $4.235 \pm 0.046$ & 2004-02-17\\
& z$^4$ & 8932 & 78.3 & $18.380 \pm 0.030$ & $3.482 \pm 0.0962$ & 2004-02-17\\
& WISE1$^6$ & 33526 & 0.818 & $14.544 \pm 0.028$ & $1.245 \pm 0.0321$ & 2014-06-07 \\
& WISE2$^6$ & 46028 & 0.242 & $13.227 \pm 0.028$ & $1.236 \pm 0.0319$ & 2014-06-07\\
& WISE3$^6$ & 115608 & 0.00652 & $10.504 \pm 0.076$ & $0.410 \pm 0.0287$ & 2014-06-07\\
& WISE4$^6$ & 220883  & 0.000509 & $8.261 \pm 0.273$ & $0.253 \pm 0.0635$ & 2014-06-07\\
\hline
\multirow{12}{*}{J1408} & fUV$^3$ & 1550 & - & - & $7.840 \pm 0.188$ & 2007-04-06\\
& u$^4$ & 3551 & 367 & $18.150 \pm 0.010$ & $20.168 \pm 0.186$ & 2004-04-13\\
& g$^4$ & 4686 & 511 & $17.843 \pm 0.015$ & $37.258 \pm 0.515$ & 2004-04-13\\
& r$^4$ & 6166 & 240 & $17.404 \pm 0.013$ & $26.129 \pm 0.314$ & 2004-04-13\\
& i$^4$ & 7480 & 128 & $17.339 \pm 0.013$ & $14.846 \pm 0.178$ & 2004-04-13\\
& z$^4$ & 8932 & 78.3 & $17.200 \pm 0.010$ & $10.322 \pm 0.095$ & 2004-04-13 \\
& J$^5$ & 12350 & 31.29 & $15.094 \pm 0.039$ & $11.361 \pm 0.837$ & 1999-02-07 \\
& H$^5$ & 16600 & 11.33 & $14.411 \pm 0.047$ & $6.226 \pm 0.918$ & 1999-02-07\\
& K$^5$ & 21900 & 42.83 & $13.620 \pm 0.039$ & $4.283 \pm 0.434$ & 1999-02-07 \\
& WISE1$^6$ & 33526 & 0.818 & $13.118 \pm 0.013$ & $4.226 \pm 0.097$ & 2014-07-07 \\
& WISE2$^6$ & 46028 & 0.242 & $12.487 \pm 0.033$& $4.291 \pm 0.087$  & 2014-07-07\\
& WISE3$^6$ & 115608 & 0.00652 & $9.527 \pm 0.063$ & $2.087 \pm 0.040$ & 2014-07-07 \\
& WISE4$^6$ & 220883 & 0.000509 & $7.829 \pm 0.486$ & $1.234 \pm 0.060$  & 2014-07-07\\ \hline

\end{tabular}
\normalsize
\\

$^1$ Units of $10^{-11}$ erg s$^{-1}$ cm$^{-2}$ \AA$^{-1}$ \\
$^2$ Units of $10^{-17}$ erg s$^{-1}$ cm$^{-2}$ \AA$^{-1}$
\\
$^3$ Data Retrieved from Galex \\
$^4$ Data Retrieved from SDSS \\
$^5$ Data Retrieved from 2MASS \\
$^6$ Data Retrieved from WISE \\
(refer to \S\ref{Phot} for further detail)

\end{table*}

\bibliographystyle{mn2e}
\bibliography{Biblio}



\end{document}